\documentclass[%
reprint,
superscriptaddress,
amsmath,amssymb,
aps,
prc
]{revtex4-2}

\usepackage{subfigure}
\usepackage{graphicx}% Include figure files
\usepackage{dcolumn}% Align table columns on decimal point
\usepackage{bm}% bold math
\usepackage{threeparttable}
\usepackage{placeins}
\usepackage{hyperref}% add hypertext capabilities
\usepackage{orcidlink}

\begin{document}

\title{Low-energy collective structure of $^{92,94}$Zr and $^{94}$Mo from ($e,e^{\prime}$) and ($p,p^{\prime}$) scattering 
\newline II. Signatures of mixed-symmetry states and origin of collectivity}

\author{C.~Walz}
\affiliation{%
Institut f\"ur Kernphysik, Technische Universit\"at Darmstadt, 64289 Darmstadt, Germany
}%
%\author{J. Carter}
\author{L.~M.~Donaldson,\orcidlink{0000-0001-8761-8257}}
\affiliation{%
iThemba Laboratory for Accelerator Based Sciences, Somerset West 7129, South Africa
}%
%\author{R. W. Fearick}
%\affiliation{%
%Department of Physics, University of Cape Town, Rondebosch 7700, South Africa
%}%
%\author{S. V. F\"ortsch}
%\affiliation{%
%iThemba Laboratory for Accelerator Based Sciences, Somerset West 7129, South Africa
%}%
%\author{H. Fujita}
%\affiliation{%
%Department of Physics, Osaka University, Toyonaka, Osaka 560-0043, Japan
%}%
%\author{C. Kremer}
\author{P.~von~Neumann-Cosel,\orcidlink{0000-0002-0256-5940}}
\email{Contact author: vnc@ikp.tu-darmstadt.de}

%\author{V.~Yu.~Ponomarev}
\author{N.~Pietralla,\orcidlink{0000-0002-4797-3032}}
\email{Email: pietralla@ikp.tu-darmstadt.de}
\affiliation{%
Institut f\"ur Kernphysik, Technische Universit\"at Darmstadt, 64289 Darmstadt, Germany
}%
%\author{R. Neveling}
%\affiliation{%
%iThemba Laboratory for Accelerator Based Sciences, Somerset West 7129, South Africa
%}%
\author{F.~D.~Smit}
\affiliation{%
iThemba Laboratory for Accelerator Based Sciences, Somerset West 7129, South Africa
}%
%\author{J.~Wambach}
%\affiliation{%
%Institut f\"ur Kernphysik, Technische Universit\"at Darmstadt, 64289 Darmstadt, Germany
%}%
%\author{C. Stahl}
%\affiliation{%
%Institut f\"ur Kernphysik, Technische Universit\"at Darmstadt, 64289 Darmstadt, Germany
%}%
%\author{J. A. Swartz}
%\affiliation{%
%iThemba Laboratory for Accelerator Based Sciences, Somerset West 7129, South Africa
%}%
%\affiliation{%
%Department of Physics, University of Stellenbosch, Matieland 7602, South Africa
%}%

\date{\today}

\begin{abstract}

This is the second of two papers discussing a new experimental signature of mixed-symmetry states (MSS) in the vibrational nuclei $^{92,94}$Zr and $^{94}$Mo based on proton and neutron transition densities.
Quasiparticle-phonon model calculations for these nuclei are extensively tested by comparison to ground-state properties, energies and moments of excited states, transition probabilities between them, and the momentum transfer dependence in $(e,e^\prime)$ and $(p,p^\prime)$ reactions. 
The overall very good agreement permits the extraction of information on the wave functions of the MSS and their fully symmetric (FSS) counterparts.
MSS can be identified by a sign change between the leading proton and neutron two-quasiparticle configurations compared to the FSS.
Possible candidates for $2^+$, $3^-$, and $4^+$ MSS are identified.
The modification of proton and neutron transition densities, which can be derived from a combined analysis of inelastic electron and proton scattering, by the sign change provide a new experimental signature of MSS.    
The collectivity of ground-state excitations of predominantly one-phonon FSS and MSS is generated to a large extent by the coupling to high-lying states forming giant resonances with the same spin and parity. 

\end{abstract}

\maketitle

\section{Introduction}
\label{sec:Introduction}

The present work is concerned with the identification and interpretation of low-energy collective states in the vibrational nuclei $^{92,94}$Zr and $^{94}$Mo with a focus on mixed-symmetry states (MSS). 
MSS are building blocks of low-energy nuclear structure with the $2^+$ state being lowest in energy in vibrational nuclei \cite{pietralla2008}.
Together with the $2^+_1$ (fully symmetric state, FSS) or other low-energy collective states like $3^-_1$, they can form multiphonon structures.
The MSS are typically identified by a strong $M1$ transition between MSS and FSS or the corresponding two-phonon states.
For spherical, vibrational nuclei they correspond  to the structures related to the scissors mode of deformed nuclei \cite{heyde2010}.
For the quadrupole case, this scheme is experimentally well established \cite{pietralla1999,pietralla2000,fransen2001} and 
serves as an experimental signature of MSS (see Refs.~\cite{stegmann2017,kern2019,kern2020,yaneva2020,stetz2025} for recent examples).
It can be interpreted within the interacting boson model 2 (IBM-2) with proton-neutron degrees of freedom \cite{arima1977} and made particularly transparent in the Q-phonon scheme \cite{pietralla1994}.

However, in many cases the situation is complicated by the mixing between one- and two-phonon states or with non-collective states of the same spin and parity leading to a fragmentation of $M1$ strength over several transitions.
Then one has to recur to microscopic models permitting a treatment of collective and non-collective features on the same footing.
In nuclei with accessible model spaces, the shell model provides a successful description of quadrupole MSS \cite{lisetskiy2000,holt2007,sieja2009}.
The quasiparticle phonon model (QPM) \cite{soloviev1992} is particularly suited for such a task, see e.g.\ Refs.~\cite{ponomarev1999,ryezayeva2002,savran2018,walz2011,burda2007}.
It permits a description of the multiphonon scheme and provides insight into the microscopic structure of the leading proton and neutron bosons.
Furthermore, because of the large single-particle model space, the coupling of valence-shell excitations to giant resonances can be studied quantitatively for tracing the formation of low-energy collectivity \cite{walz2011}.

The purpose of this paper is threefold.
In Sec.~\ref{sec:QPM Calculations}, an extensive comparison with ground-state (g.s.) properties, excitation spectra, electromagnetic moments and transition strengths, as well as the proton and electron scattering data presented in Paper I is made for the nuclei of interest. 
The successful description, especially for $2^+$ states, permits the extraction of the underlying wave functions.
Besides the identification of one-phonon $2^+$ MSS and their features, candidates for predominantly one-phonon $3^-$ and $4^+$ MSS are discussed. 
In the terminology of the IBM-2, they correspond to a mixed-symmetric coupling of $f$-bosons or $g$-bosons to the $sd$-boson space, respectively.  While the $sdf$-IBM-2 has been formulated in Ref.~\cite{smirnova2000}, a corresponding formulation of the $sdfg$-IBM-2 is still lacking.
The identification of one-phonon 3- and 4+ MSS is complicated by the fact that two-phonon configurations with MS character in the $sd$-boson space with the same spin and parity quantum numbers may occur at the comparable excitation energies. 
Large $M1$ transition strengths may then not suffice for a unique identification, as will be discussed in detail below. 
The QPM results may then be particularly useful to distinguish predominantly one-phonon from two-phonon excitations.  

Alternative signatures for the identification of one-phonon MSS based on proton and neutron transition matrix elements and transition densities are then tested in Sec.~\ref{sec: Signatures of 2$^+$ mixed-symmetry states}. 
The latter, which can be decomposed in a combined analysis of $(e,e')$ and $(p,p')$ scattering, provide an independent way to identify the MSS based on the different momentum transfer dependence compared to all other $2^+$ states \cite{burda2007}. 
Finally, in Sec.~\ref{sec:IV}, the importance of high-lying two-quasiparticle (2qp) states forming giant resonances of the same spin and parity for the generation of collectivity is discussed for $2^+$, $3^-$, and $4^+$ FSS and MSS candidates. 

\section{Comparison with QPM Calculations}
\label{sec:QPM Calculations}

This section presents an extensive comparison of QPM predictions for the low-energy states in $^{92,94}$Zr and $^{94}$Mo \cite{ponomarevpc} with experimental data.
The results demonstrate the predictive power of the model, which allows in turn to draw conclusions on a new experimental signature of quadrupole MSS based on transition radii and the possible existence of octupole and hexadecapole MSS.
QPM calculations based on subsets of the data have been presented for $2^+$ MSS in $^{92}$Zr \cite{loiudice2004,loiudice2006} and $^{94}$Mo \cite{loiudice2000} with similar results.

\subsection{Details of the model}
\label{subsec: Details of the QPM model}

The QPM approach applied in the present work is described in detail in Ref.~\cite{walz2014}.
A Woods-Saxon potential, ﬁxed to the properties of neighboring nuclei, was used to obtain the single-particle basis. 
The strength of the pairing force was ﬁtted to odd-even mass differences, and the strength of the residual interaction was fixed to describe the $B(E2)$ values and excitation energies of the $2_1^+$ state. 
Numerical values for the Woods-Saxon potentials, the pairing strength and the parameters of the residual interactions are given in Tab.~5.1 of Ref.~\cite{walz2014}. 
No additional parameters are needed for the coupling to multiphonon states.

Since the QPM (in contrast to the shell model) possesses a single-particle basis sufficiently large to fulﬁll the energy-weighted sum rules, no effective charges are necessary to describe the $B(E2)$ strengths. 
$M1$ transitions were calculated
assuming a spin quenching factor $g_s = 0.6$.
Calculations of inelastic scattering cross sections were performed using QPM transition densities as input.
Proton scattering was studied with the distorted-wave Born approximation (DWBA) code DWBA07 \cite{dwba07} using the Love-Franey proton-nucleus effective interaction \cite{love1981,franey1985}.
Electron scattering calculations utilized the DWBA code of Heisenberg \cite{heisenberg1983}.

\subsection{Test of the ground-state properties}
\label{subsec: Ground state properties}

\begin{figure}[b] 
\centering
    \includegraphics[width=0.8\columnwidth]{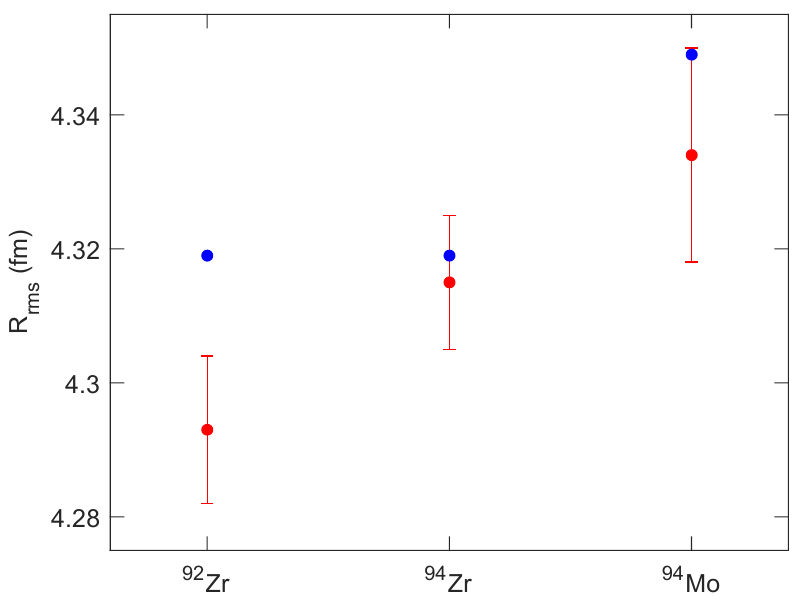}
    \caption[Comparison of the calculated and measured root-mean-square radii of the charge distributions.]{\label{Fig: elastic electron}
    Experimental (red circles) root-mean-square radii of the charge distributions \cite{devries1987} in comparison with predictions of the QPM approach described in the text (blue circles).}
\end{figure}

Since the one-phonon states are created by acting on the ground state, it is important to verify the predictions of the QPM for the composition of the ground-state wave function against experimental observables. 
Figure~\ref{Fig: elastic electron} compares the root-mean-square radii of the proton (or charge) density distributions calculated in the QPM to experimental values obtained from elastic electron scattering for the nuclei $^{92,94}$Zr and $^{94}$Mo \cite{devries1987}. 
The QPM slightly overestimates the proton radius of $^{92}$Zr by 0.03~fm, while the predicted radii of $^{94}$Zr and $^{94}$Mo are in excellent agreement with data within error bars suggesting that the charge density distributions are correctly described.

\begin{figure}
\centering
        \includegraphics[width=\columnwidth]{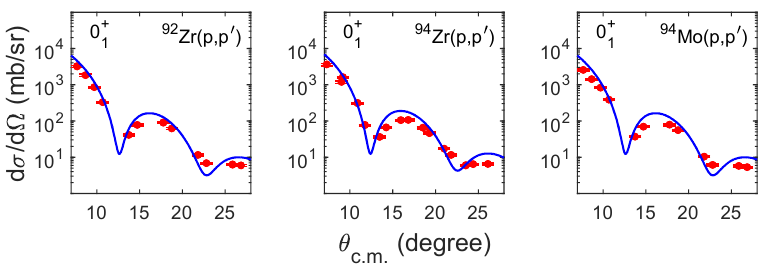}
    \caption[Comparison of the calculated and measured elastic proton scattering cross sections.]{ \label{Fig: elastic proton}
    Comparison of elastic proton scattering cross sections off $^{92,94}$Zr and $^{94}$Mo at an energy of 200~MeV with QPM predictions.}
\end{figure}

In the proton scattering experiments presented in Paper I, elastic-scattering cross sections were also measured.
They are used to test the optical model results obtained by folding the proton and neutron ground-state densities with the effective QPM interaction. 
The comparison between theory and data is displayed in Fig.~\ref{Fig: elastic proton}. 
Despite some deviations at small scattering angles, the general description of the data is successful in terms of shapes as well as magnitudes. 
It should be noted that the calculated cross sections are absolute and no changes were made to the effective nucleon-nucleus interaction. 
This is also true for all other theoretical proton-scattering cross sections shown in this work.

The fair description of charge radii and elastic proton scattering cross sections validate the QPM ground-state wave functions as well as the effective nucleon-nucleus interaction. 
This places all calculations of the properties of excited states presented in the following sections on a solid footing.

\subsection{Description of $2^+$ states}
\label{subsec: 2+ states}

\subsubsection{Energies and transition strengths}
\label{subsec: 2+ Energies and transition strength}

\begin{table} 
\centering
\caption[Structure of the QPM wave function of the 2$^+$ states in terms of QRPA phonons.]
{\label{tab: quadrupole QPM wave functions} Excitation energies and leading components of the QPM wave functions of the three lowest quadrupole states in the nuclei $^{92,94}$Zr and $^{94}$Mo.}
\begin{tabular}{cccccc}
\hline
\hline
Nucleus &                  & \multicolumn{2}{c}{$E_{\rm x}$ (keV)} & \multicolumn{2}{c}{Structure} \\
             &                         & Exp.          & QPM &  &  \\
\hline
           & 2$_1^+$  & 934  & 1025  & 91$\%[2^+_1]$   & \\
$^{92}$Zr  & 2$_2^+$  & 1847 & 1983  & 91$\%[2^+_2]$   & \\
           & 2$_3^+$  & 2066 & 2043  & 17$\%[2^+_4]$ + 13$\%[2^+_5]$ - 54$\%[2^+_1 \otimes 2^+_1 ]$  & \\
\hline
           & 2$_1^+$  & 919  &  856  &  93$\%[2^+_1]$  & \\
$^{94}$Zr  & 2$_2^+$  & 1671 & 1961  &  35$\%[2^+_2]$ + 44$\%[2^+_1 \otimes 2^+_1]$   & \\
           & 2$_3^+$  & 2366 & 2090  &  49$\%[2^+_2]$ + 31$\%[2^+_1 \otimes 2^+_1]$   & \\
\hline
          & 2$_1^+$  & 872  &  822  & 86$\%[2^+_1]$     & \\
$^{94}$Mo & 2$_2^+$  & 1864 & 1470  & 13$\%[2^+_3]$ + 63$\%[2^+_1 \otimes2^+_1]$     & \\
          & 2$_3^+$  & 2067 & 1870  & 75$\%[2^+_2]$ - 10$\%[2^+_1 \otimes4^+_1]$     & \\   
\hline
\hline
\end{tabular}
\end{table}

Table~\ref{tab: quadrupole QPM wave functions} compares the experimental excitation energies of the lowest three $2^+$ states to the QPM results
and provides the dominant contributions of QRPA solutions to the QPM wave functions. 
The leading components of the two lowest QRPA phonons are given in Table \ref{tab: quadrupole QRPA wave functions}.

\begin{table}
\centering
\caption[Structure of the quadrupole QRPA-phonons in terms of two-quasiparticle states.]{\label{tab: quadrupole QRPA wave functions} 
Leading components of the wave functions of the [2$^+_1]$- and [2$^+_2]$ QRPA phonons for the nuclei $^{94}$Mo and $^{92,94}$Zr. $\psi$ and $\phi$ are the forward and backward amplitudes defined e.g.\ in Eq.~(17) of Ref.~\cite{vonneumanncosel2019}. 
The percentage numbers are the contributions of the corresponding 2qp states to the norm of the wave functions. }
\begin{tabular}{cccccccc}
\hline
\hline
            Nucleus &  QRPA & \multicolumn{3}{c}{($2d_{5/2}\otimes 2d_{5/2})_n$}  & \multicolumn{3}{c}{($1g_{9/2}\otimes 1g_{9/2})_p$}  \\
  &  phonons        &$\psi$ &$\phi$ & $\%$  & $X$   & $Y$ & $\%$  \\
\hline
$^{94}$Mo    & [2$^+_1$]  &  1.06     & 0.20    & 54.64   & 0.87      & 0.30        & 32.95   \\
             & [2$^+_2$]  &  -0.95    & 0.09    & 44.49   & 0.96      & 0.08        & 45.87   \\
\hline
$^{92}$Zr    & $[2^+_1]$  &  1.20     & 0.16    & 70.98   & 0.51      &  0.19       & 11.03   \\
             & $[2^+_2]$  &  -0.76    & 0.11    & 28.30   & 0.80      &  0.09       & 31.66   \\
\hline
$^{94}$Zr    & $[2^+_1]$  &  1.27     & 0.20    & 78.20   & 0.42      & 0.20        & 6.93    \\
             & $[2^+_2]$  &  -0.65    & 0.12    & 20.27   & 0.64      & 0.10        & 19.83   \\
\hline
\hline
\end{tabular}
\end{table}

In $^{92}$Zr and $^{94}$Zr, fair agreement is observed with exception of the $2_2^+$ state energy in $^{94}$Mo, which is $\sim$400~keV lower than the experimental value. 
The unperturbed energy of its main $[2^+_1 \otimes2^+_1]$ component lies at 2618~keV. 
It is pushed down in energy by the interaction with three-phonon states at higher energies. 
The very low energy of the final state (1470~keV) indicates that the interaction strength between this two-phonon state and the three-phonon states is overestimated by the model.

In $^{94}$Zr, the QPM calculations exhibit additional problems when describing the excitation energies as well as the electromagnetic transition strengths. 
Due to proximity of the unperturbed $[2^+_1 \otimes2^+_1]$ and $[2^+_2]$, the second and the third 2$^+$ state have large two-phonon components and, therefore, enhanced $B(E2)$-values compared to the 2$^+_1$ state.
This is not seen in experiment, where the transition of the 2$^+_2$ state to the 2$^+_1$ state is $0.061^{+0.13}_{-0.06}$~W.u.\ only.
However, a slightly larger separation of the two unperturbed energies by a few hundred keV would drastically reduce the mixing and solve the discrepancy. 
It might be possible to achieve this additional separation by additionally taking into account the quadrupole particle-particle interaction not considered in the present calculations. 
Due to this problem, the interaction between different phonons is neglected in $^{94}$Zr for the 2$^+$ states, i.e., further discussion of the experimental data is restricted to the QRPA results. 
In order the achieve a reasonable description of the $B(E2)$ values in $^{94}$Zr, it was necessary to lower the energy of the $(2d_{5/2}\otimes 2d_{5/2})_n$ 2qp state artificially by 200~keV. 
Without this change, the QPM predicts a $B(E2)$ value for the MSS that is a factor of four too low. 
This highlights that MSS are sensitive probes of the underlying shell structure. 

\begin{table} \centering
\caption[Comparison of the calculated and measured transition strengths of 2$^+$ states.]
{\label{tab: quadrupole transition matrix elements} 
Experimental transition strengths of the decay of the three lowest $2^+$ states to the g.s.\ and between them in $^{92}$Zr \cite{fransen2005}, $^{94}$Zr \cite{elhami2008}, and $^{94}$Mo \cite{fransen2003} in comparison with QPM predictions. QPM$_{\rm{p,n}}$ denote the electric quadrupole transition strengths  caused by the proton and neutron parts of the wave functions, respectively.
The theoretical results for $^{94}$Zr are on the QRPA level only.}
\begin{threeparttable}[c]
\begin{tabular}{ccccccc}
\hline
\hline
              &                             & \multicolumn{3}{c}{$B(E2)$(W.u.)} & \multicolumn{2}{c}{$B(M1)(\mu^2_N$)} \\
Nucleus   &  $J_i\rightarrow J_f$  & Exp.       & QPM$_{\rm p}$        & QPM$_{\rm n}$ & Exp.      & QPM \\
\hline
          & 2$_1^+$ $\rightarrow$ 0$_1^+$  & 6.5(5)               & 5.9           &  14.1      &          &     \\
          & 2$_2^+$ $\rightarrow$ 0$_1^+$  & 3.5(4)               & 2.7           &   3.7      &          &     \\
$^{92}$Zr & 2$_3^+$ $\rightarrow$ 0$_1^+$  & $<$0.005               & 0.1           &   0.1      &          &     \\
          & 2$_2^+$ $\rightarrow$ 2$_1^+$  & 0.4$^{+0.5}_{-0.3}$  & 0.2           &            &  0.37(4) & 0.52\\
          & 2$_3^+$ $\rightarrow$ 2$_1^+$  & $<$16                & 6.4           &            & $<$0.024 & 0.02\\
\hline
          & 2$_1^+$ $\rightarrow$ 0$_1^+$  & 4.9(11)                   &   4.9  & 15.0 &                &     \\
          & 2$_2^+$ $\rightarrow$ 0$_1^+$  & 3.9(3)                    &   3.4  &  6.0 &                &     \\
$^{94}$Zr & 2$_2^+$ $\rightarrow$ 2$_1^+$  & 0.061$^{+0.13}_{-0.06}$   &   0.005 &   & 0.085$^{+6}_{-7}$ & 0.40\\
          & 2$_3^+$ $\rightarrow$ 0$_1^+$  & 0.019$^{+0.011}_{-0.012}$ &   0.005  &  &                &     \\
          & 2$_3^+$ $\rightarrow$ 2$_1^+$  & 60$^{+24}_{-30}$          &   49.1  &   &                &     \\
\hline
          & 2$_1^+$ $\rightarrow$ 0$_1^+$  & 16.0(2)              &  12.8         & 24.7          &          &   \\
          & 2$_2^+$ $\rightarrow$ 0$_1^+$  & 0.33(11)             &  0.01         & 0.01          &          &   \\
$^{94}$Mo & 2$_3^+$ $\rightarrow$ 0$_1^+$  & 2.2(2)               &  1.54         & 1.66          &          &   \\
          & 2$_2^+$ $\rightarrow$ 2$_1^+$  & 60$^{+20}_{-30}$     &  17.4         &            & 0.026$^{+0.041}_{-0.016}$ & 0.008   \\
          & 2$_3^+$ $\rightarrow$ 2$_1^+$  & 4.9$^{+3.0}_{-2.3}$  &  0.3          &               & 0.56(5)  & 0.48 \\
\hline
\hline
\end{tabular}
%  \begin{tablenotes}\footnotesize
%     \item[1] QRPA values
%  \end{tablenotes}
\end{threeparttable}
\end{table}

Table~\ref{tab: quadrupole transition matrix elements} summarizes the experimental information on $E2$ and $M1$ transitions between the states of interest.
For $E2$ transitions, these can be compared to the QPM predictions, denoted QPM$_{\rm p}$ (since only protons contribute to the electromagnetic transition strengths).
The corresponding neutron strengths, denoted QPM$_{\rm n}$, are also given.
They contribute in the theoretical description of $(p,p^\prime)$ scattering cross sections sensitive to the matter distribution. 
The main signature of a one-phonon quadrupole MSS is an enhanced $B(M1)$ value to the FSS quadrupole state. 
As can be seen in Tab.~\ref{tab: quadrupole transition matrix elements}, this holds for the 2$_2^+$ states of $^{92,94}$Zr and for the 2$_3^+$ state of $^{94}$Mo. 
Hence, they were identified in the literature  as the states carrying the main fragment of the mixed-symmetry states \cite{pietralla1999, fransen2003, fransen2005, elhami2008}. 
This view is supported by the QPM calculations capable of reproducing the enhanced $B(M1)$ values.
The restriction of calculations to the QRPA level for $^{94}$Zr leads to an overestimate of the experimental value by a factor of about five. 

The mechanism behind the large $M1$ transition strengths can be understood as follows: 
The anomalous proton and neutron $g$ factors as well as the main proton and neutron 2qp states in the wave function of the MSS have different signs. 
Therefore, the  $M1$ matrix elements of the main 2qp configurations with opposite phases add up coherently.  
In Sec.~\ref{subsec: Proton-neutron transition radii}, an alternative signature is discussed, which proves the mixed-symmetric character of the corresponding states in $^{92,94}$Zr and $^{94}$Mo independently of electromagnetic transition strengths.

In addition to energies and electromagnetic transition strengths, $g$ factors are useful observables to test the QPM predictions.
They provide independent information on the proton-neutron content of the wave functions. 
The $g$ factors of all three $2_1^+$ states and of the MSS of $^{92,94}$Zr are shown in Tab.~\ref{tab: quadrupole g-factors}. 
For the 2$^+_3$ state in $^{94}$Mo, no $g$ factor is known. 
The QPM results for the $g$ factors are in good agreement with data, especially regarding the signs, which are correctly described. 
This gives additional support for the QPM wave functions in Tab.~\ref{tab: quadrupole QRPA wave functions} and the simple two-state model discussed below. 
Both models predict a neutron dominance of the 2$^+_1$ state of $^{92,94}$Zr due to the energy gap between the lowest proton and neutron 2qp states. 
The negative signs of the $g$ factors confirm this prediction. 
Due to the reduced energy difference, no strong neutron dominance is expected for $^{94}$Mo. 
Again, the sign of the $g$ factor of the $2_1^+$ state supports this assumption.

\begin{table}
\centering
\caption[Comparison of the calculated and measured g-factors of 2$^+$ states.]{\label{tab: quadrupole g-factors}
Experimental $g$ factors of the lowest $2^+$ states in $^{92,94}$Zr \cite{werner2008} and $^{94}$Mo \cite{mantica2001} in comparison with QPM predictions.
The theoretical results for $^{94}$Zr are on the QRPA level only.}
\begin{threeparttable}[c]
\begin{tabular}{cccc}
\hline
\hline
Nucleus  &  $g(J^{\pi}$)($\mu_n$)      & Exp.       & QPM     \\
\hline
$^{92}$Zr     &  $g(2^+_1$)                 & -0.18(1)  & -0.09   \\
              &  $g(2^+_2$)                 & 0.76(50)  & 0.73    \\
\hline
$^{94}$Zr     &  $g(2^+_1$)                 & -0.32(2)  & -0.13   \\
              &  $g(2^+_2$)                 & +0.88(27) &  0.38  \\
\hline
$^{94}$Mo     &  $g(2^+_1$)                 & 0.275(75) & 0.44    \\
\hline
\hline
\end{tabular}
\end{threeparttable}
\end{table}

\subsubsection{Proton scattering}
\label{subsubsec: 2+ proton scattering}

The $B(M1)$ values in Tab.~\ref{tab: quadrupole transition matrix elements} and $g$ factors in Tab.~\ref{tab: quadrupole g-factors} are
sensitive to the main components of the neutron and proton parts of the wave functions.
On the other hand, absolute $E2$ transition matrix elements to the ground state are mainly determined by a multitude of small contributions from high-lying 2qp configurations that may provide 
large transition matrix elements, each, but contribute individually with small amplitudes to the wave functions (cf.\ Ref.~\cite{walz2011} and Sec.~\ref{sec:IV}). 
The correct description of $B(E2)$ values proves that the QPM is capable of accounting for the collectivity of the proton components. 
The absolute proton scattering cross sections measured in this work permit a test of whether the same holds for the neutron parts of the QPM wave functions.

\begin{figure}
\centering
    \includegraphics[width=\columnwidth]{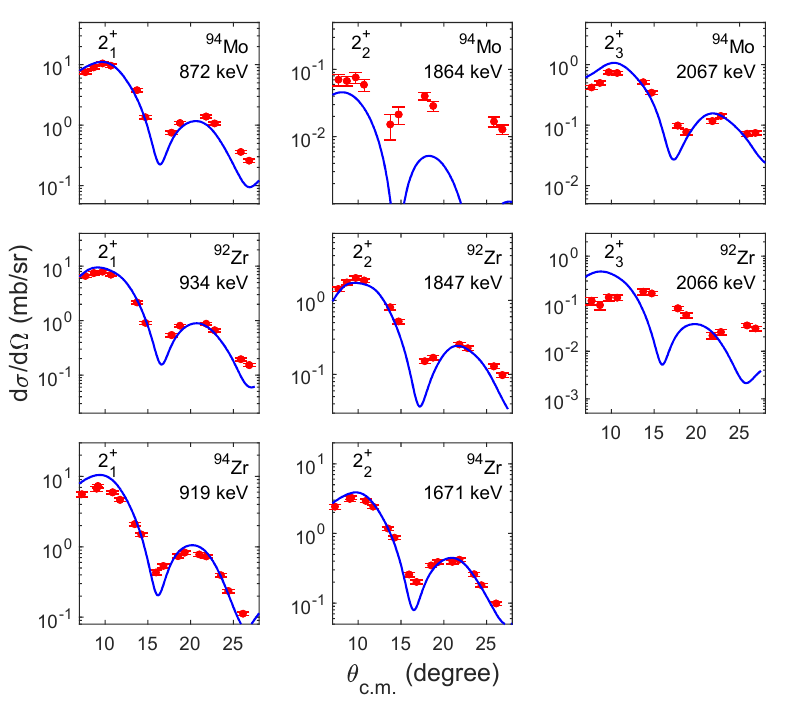}
    \caption[Comparison of the calculated and measured proton scattering cross sections of 2$^+$ states.]
    {\label{Fig: pp comparison of quadrupole states}
    Comparison of the proton scattering cross sections populating the first three quadrupole states in $^{92,94}$Zr and $^{94}$Mo to QPM calculations assuming one-step excitations (blue lines). 
    The calculations for $^{94}$Zr are on the
    QRPA level only.}
\end{figure}

Figure~\ref{Fig: pp comparison of quadrupole states} compares the theoretical cross sections (blue curves) with the experimental data for the transitions to the lowest three 2$^+$ states.
Data for the 2$_3^+$ state of $^{94}$Zr are left out.
It is expected to be predominantly a member of the two-phonon triplet due to its large $B(E2)$ value to the 2$^+_1$ state. Accordingly, the direct excitation probability in the $(p,p^\prime)$ reaction is small. 
Furthermore, the state could not be fully resolved from a close-lying $4^+$ state, cf.\ Paper I.
The description of the cross sections of the 2$_1^+$ and 2$_{\rm ms}^+$ states is excellent in terms of magnitudes as well as of shapes. 
Proton scattering at 200 MeV is sensitive to both proton and neutron transition matrix elements. 
Since the $B(E2)$ values of the QPM are in agreement with experiment, the correct description of the proton scattering cross sections indicates that the neutron transition matrix elements of the QPM are also reasonable.

The DWBA calculations fail to describe the cross section of the 2$_2^+$ of $^{94}$Mo and 2$_3^+$ state of $^{92}$Zr.
The calculations consider direct excitations from the ground state only and neglect coupled-channel effects. Both states are predicted to contain large two-phonon components $[2^+_1 \otimes 2^+_1]$, see Tab.~\ref{tab: quadrupole QPM wave functions}.
Coupling to the 2$^+_1$ state is then expected to play a crucial role in the description of the cross sections. %It is not possible to perform coupled-channel calculations with the computer code DWBA07~\cite{dwba07} used here. 
In Refs.~\cite{burda2007,walz2010}, some attempts were made to describe the angular distribution of the cross sections with the CHUCK3~\cite{chuck3} code, which is able to include multistep excitations. 
However, the achieved description was not very good, and 
the question of whether the unusual angular distributions of both states -- which are indeed a strong hint for a dominant two-phonon component~\cite{deleo1998} -- can be explained by the inclusion of multistep processes remains open.

\subsubsection{Electron scattering}
\label{subsubsec: 2+ Electron scattering}

\begin{figure}
\centering
    \includegraphics[width=0.67\columnwidth]{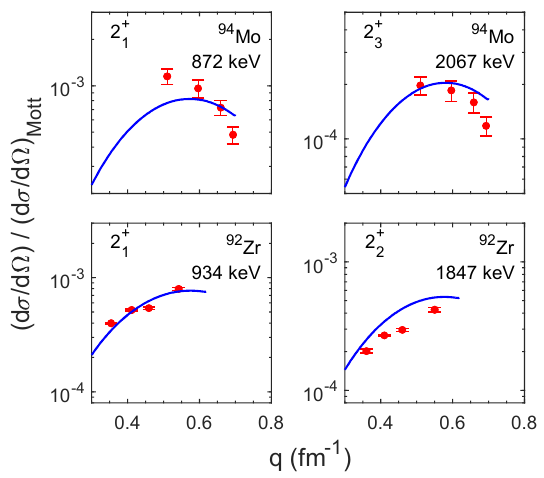}
    \caption[Comparison of the calculated and measured electron scattering cross sections of 2$^+$ states.]
     {\label{Fig: ee comparison of quadrupole states}
    Comparison of electron scattering cross sections of the $2^+$ FSS and MSS in $^{94}$Mo \cite{burda2007} and $^{92}$Zr \cite{scheikh2013} with QPM predictions (blue lines). 
    }
\end{figure}

Additional information on the proton transition densities can be obtained from electron scattering data. Electron scattering
at low momentum transfers mainly probes the $B(E2)$ values and the proton transition radii. 
The electron scattering form factors calculated in the framework of the QPM are compared to the data of Refs.~\cite{burda2007, scheikh2013} in Fig.~\ref{Fig: ee comparison of quadrupole states}. 
In $^{92}$Zr, an acceptable description of the experimental form factors is achieved. 
On the other hand, in $^{94}$Mo, the theoretical form factor is systematically shifted to higher momentum transfers with respect to the experimental one. 
This points to an underestimation of the proton transition radii.
Attempts to fix the problem by slightly changing parameters of the Wood-Saxon potential were unsuccessful.

\subsubsection{Two-state model}
\label{subsubsec: Two-state model}

\begin{figure} 
\centering
    \includegraphics[width=0.6\columnwidth]{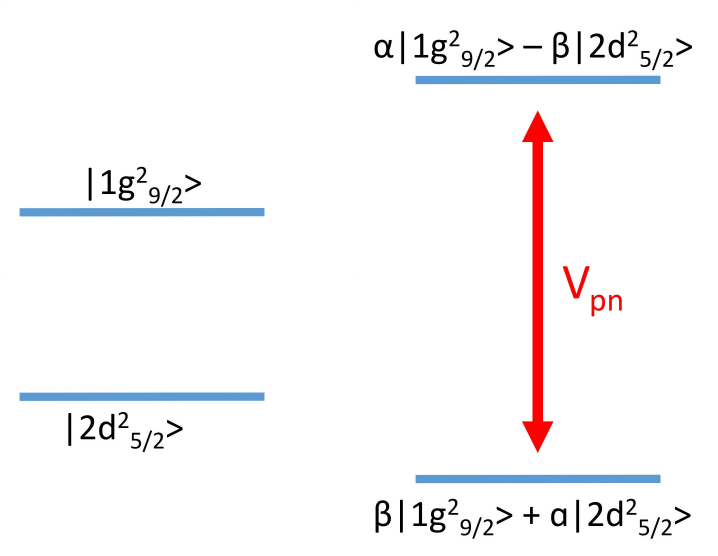}
    \caption[Two-state mixing model.]
     {\label{Fig: Two-state mixing scheme}
    Two-state mixing model to explain the structure of the leading components of the QRPA phonons in Tab.~\ref{tab: quadrupole QRPA wave functions}. 
    The residual proton-neutron interaction $V_{pn}$ mixes the unperturbed 2qp states.}
\end{figure}

Table~\ref{tab: quadrupole QRPA wave functions} displays the two leading components of the wave functions of the first and second quadrupole QRPA phonons in $^{94}$Mo and $^{92,94}$Zr. The same two 2qp states are important in all states.
The neutron ($2d_{5/2}\otimes 2d_{5/2})_n$ and the proton ($1g_{9/2}\otimes 1g_{9/2})_p$ 2qp states are in phase for the $[2^+_1]$ phonons and out of phase for the $[2^+_2]$ phonons (the forward RPA amplitude $X$ determines the sign).
The formation mechanism of this structure can be understood using a simple two-state model. 
On the left side of Fig.~\ref{Fig: Two-state mixing scheme}, the two unperturbed 2qp states are schematically shown. 
If one takes the proton-neutron residual interaction into consideration, both 2qp states mix, forming two states with different relative phase between the proton and neutron components.
The strength of the residual proton-neutron interaction and the energy difference of the unperturbed 2qp states determine the degree of mixing, i.e., the amplitudes $\alpha$ and $\beta$. 
In case of strong mixing, $\alpha\approx\beta\approx1/\sqrt{2}$. 

The energy differences of the pure (2d$_{5/2} \otimes$ 2d$_{5/2})_n$ and ($1g_{9/2}\otimes 1g_{9/2})_p$ states are 781 keV in $^{94}$Mo, 1255 keV in $^{92}$Zr and 1578 keV in $^{94}$Zr. 
The strength of the residual interaction is comparable in all three nuclei. 
Therefore, one  expects that the degree of mixing increases from $^{94}$Zr to $^{94}$Mo. 
This is confirmed by the wave function components in Tab.~\ref{tab: quadrupole QRPA wave functions}. 
The $[2^+_1]$ phonons in $^{94}$Zr and $^{92}$Zr are neutron dominated ($^{92}$Zr to a lesser extent), while the 2$_1^+$ state in $^{94}$Mo is close to the case $\alpha = \beta$. 
This phenomenon has previously been discussed in the literature in terms of 'configurational isospin polarization' \cite{holt2007}.

In general, many other 2qp states contribute to the wave functions in the QRPA calculations weakening such a two-state mixing picture. 
A large fraction of the wave function, however, can be understood in this simple scheme. 
Heyde and Sau \cite{heyde1986} demonstrated that, independent of the particular model, the appearance of FSS and MSS quadrupole states is a general feature of the nucleus due to its proton-neutron two-component character.
It works particularly well in the three nuclei investigated here because the relevant proton and neutron 2qp states are slightly separated in energy from other 2qp states. 

%\input{biblio_experimentalResults.tex}

%%%%%%%%%%%%%%%%%%%%%%%Intersection of two papers%%%%%%%%%%%%%%%%%%%%%%%%%%%%%

\subsection{ Description of $3^-$ and $5^-$ states}
\label{subsec: 3- and 5- states}

\begin{table}[b]
\centering
\caption[Structure of the octupole QRPA-phonons in terms of two-quasiparticle states.]
{\label{tab: octupole QRPA wave functions} 
Same as Tab.~\ref{tab: quadrupole QRPA wave functions} but for octupole QRPA phonons.}
\begin{tabular}{cccccccc}
\hline
\hline
             & & \multicolumn{3}{c}{($2d_{5/2}\otimes1h_{11/2})_n$}  & \multicolumn{3}{c}{(2p$_{3/2}\otimes1g_{9/2})_p$} \\
$^{\rm A}X$  & QRPA phonons        &$\psi$ &$\phi$ & $\%$  & $X$   & $Y$ \\
%$  \\
\hline
$^{94}$Mo    & $[3^-_1]$  & 0.44      & 0.13    & 18.01   & 0.73      & 0.18 & 49.87  \\
             & $[3^-_3]$  & -0.84     & 0.00    & 71.72   & 0.47      & -0.03  & 21.89  \\
\hline
$^{92}$Zr    & $[3^-_1]$  & 0.38      & 0.12    & 13.40   & 0.80      & 0.17 & 60.39  \\
             & $[3^-_3]$  & -0.89     & 0.00    & 78.35   & 0.41      & -0.04   & 16.56  \\
\hline
$^{94}$Zr    & $[3^-_1]$  & 0.58      & 0.20    & 29.59   & 0.72      & 0.21 & 46.99 \\
             & $[3^-_3]$  & -0.79     & 0.01    & 62.26   & 0.54      & -0.03  & 29.27 \\
\hline
\hline
\end{tabular}
\end{table}

This section examines the properties of the 3$^-$ and 5$^-$ states of $^{92,94}$Zr and $^{94}$Mo.
Due to the negative parity, different 2qp states are important for the $3^-,5^-$ states compared to 2$^+,4^+$ states. 
In order to form low-lying one-phonon symmetric and mixed-symmetric states, both proton and neutron 2qp states have to be present at low energies. 
For the negative-parity states, this can only be achieved when unique-parity single-particle states contribute to the two main components, because cross-shell excitations are too high in energy to contribute.
In the investigated nuclei, these are $1g_{9/2}$ and $1h_{11/2}$ for protons and neutrons, respectively. 
Indeed, both contribute to the two dominant 2qp components, ($2d_{5/2}\otimes1h_{11/2})_n$ and ($2p_{3/2}\otimes1g_{9/2})_p$, which determine the structure of the $[3^-_1]$ and $[3^-_3]$ phonons, see Tab.~\ref{tab: octupole QRPA wave functions}).
Like for the 2$^+$ states, the two main 2qp components are in phase for the $[3^-_1]$-phonon and out of phase for the $[3^-_3]$ phonon, i.e., the QPM predicts an octupole MSS. 
However, the formation of a clean one-phonon $3^-$ MSS is hindered by the occurrence of a $3^-$ state originating from the $2^+ \otimes 3^-$ two-phonon coupling at almost the same energy as expected for the one-phonon $3^-$ MSS. 
Indeed, this two-phonon configuration mixes into the wave functions of the two lowest-energy $3^-$ states of $^{92}$Zr and $^{94}$Mo
The $[3^-_2]$ phonons are predicted to be pure 2qp states for all three nuclei and are, thus, not considered here.

No MSS are predicted for $5^-$ states.
In all three nuclei, the $[5^-_1]$ phonons are pure ($2p_{1/2}\otimes 1g_{9/2})_p$ 2qp states  with amplitudes larger than 90\%. 
None of the 5$^-$ QRPA phonons exhibit a structure compatible with the two-state picture. 
This is likely due to the large energy difference of about 1.6 MeV between the lowest proton and neutron 2qp states and the weakness of the residual proton-neutron interaction. 

\subsubsection{Energies and transition strengths}
\label{subsubsec: 3- and 5- energies and transition strengths}

\begin{table} 
\centering
\caption[Structure of octupole and 5$^-$ QPM wave functions in terms of QRPA-phonons.]{\label{tab: octupole QPM wave functions}
Same as Tab.~\ref{tab: quadrupole QPM wave functions} but for $3^-$ and $5^-$ QPM states.}
\begin{tabular}{cccccc}
\hline
\hline
  &  State                  & \multicolumn{2}{c}{$E$ (keV)} & \multicolumn{2}{c}{Structure} \\
             &                         & Exp.         & QPM &  &  \\
\hline
             &  3$_1^-$  & 2534 & 2429  & 86\% $[3^-_1]$ + 10\% $[2^+_1\otimes3^-_1]$   & \\
$^{94}$Mo    &  3$_2^-$  & 3012 & 3238  & 10\% $[3^-_1]$ + 11\% $[3^-_3]$ - 57\% $[2^+_1\otimes3^-_1]$  & \\
             &  5$_1^-$  & 2611 & 2422  & 92\% $[5^-_1]$     & \\
\hline
             & 3$_1^-$  & 2340 & 2342  & 88\% $[3^-_1]$ + 11\% $[2^+_1\otimes3^-_1]$     & \\
$^{92}$Zr    & 3$_2^-$  & 3446 & 3766  & 7\% $[3^-_1]$ - 12\% $[3^-_3]$ - 55\% $[2^+_1\otimes3^-_1]$ & \\
             & 5$_1^-$  & 2486 & 2432  & 94\% $[5^-_1]$   & \\
\hline
\hline
\end{tabular}
\end{table}

The predicted excitation energies and the full wave functions of the QPM states are shown in Tab.~\ref{tab: octupole QPM wave functions}. 
The comparison with experimental electromagnetic transition strengths is given in Tab.~\ref{tab: octupole transition matrix elements}. 
The agreement with the experimental excitation energies of the lowest $3^-$ and $5^-$ states shown is good, except for the $3^-_2$ state of $^{92}$Zr with a difference of 330 keV.
Although the 3$^-_1$ states of $^{92}$Zr and $^{94}$Mo are dominated by the $[3^-_1]$ phonons, $[2^+_1\otimes3^-_1]$ components contribute with significant amplitudes. 
The $3^-_2$ states are mainly members of the $[2^-_1 \otimes 3^-_1]$ quintuplets. 
The QPM calculation reproduces the predominant two-phonon character of the $3^-_2$ state with unusual positive anharmonicity, which was previously reported for the corresponding $1^-$ state of $^{92}$Zr \cite{fransen2004}, too. 

The two-phonon states are expected to decay with large $B(E2)$ values to the 3$^-_1$ states. 
An upper limit of 35 W.u.\ was derived for this transition in $^{94}$Mo \cite{scheck2010}, which is not at variance with the predicted structure. 
The $B(E3)$ transition strengths to the $3^-_1$ states are reproduced well for all three nuclei.
Reference~\cite{scheck2010} also claims the 3$^-_2$ state of $^{94}$Mo to be a candidate for the octupole MSS based on the observation of a strong $M1$ transition to the $3^-_1$ state. 
The QPM calculations do not contradict these results, since the $[3^-_3]$ phonons contribute with amplitudes of $\sim$10$\%$ to the 3$^-_2$ states of $^{92}$Zr and $^{94}$Mo.
The unperturbed energies of the $[3^-_3]$ phonons -- and hence also the main fragments -- are expected at excitation energies well above 4 MeV for all three nuclei.
In the full QPM calculation, the high energies of the theoretical $3^-_3$ states remain and can be seen as a shell structure effect due to the absence of 2qp states below 3 MeV.

As in the case of the $4^+$ states discussed below, a large $B(M1)$ value can be a strong indication for a MSS. 
This holds when the $M1$ transition is caused by a fragmentation of underlying proton and neutron configurations with opposite phases. 
This is precisely the microscopic mechanism from which MSS are formed. 
However, the QPM calculations imply that the one-phonon octupole MSS may be quite fragmented. 
Then, similar to the case of the 4$^+$ states, a large $B(M1)$ value is not a unique signature of an MSS. 
It can also be caused e.g.\ by  a fragmentation of the $[3^-_1]$ phonon and a transition between these amplitudes in different $3^-$ states. 
The theoretical strength of this particular transition depends upon the degree of mixing of the two main 2qp states and the assumed spin-quenching factor.

\begin{table} 
\centering
\renewcommand*{\arraystretch}{1.3}
\caption[Comparison of the calculated and measured transition strengths of octupole and 5$^-$ states.]
{\label{tab: octupole transition matrix elements} Same as Tab.~\ref{tab: quadrupole transition matrix elements} but for $3^-$ and 5$^-$ states.}
\begin{threeparttable}[c]
\begin{tabular}{ccccccccc}
\hline
\hline
              &                             & \multicolumn{4}{c}{$B(EJ)$(W.u.)} & \multicolumn{3}{c}{$B(M1)(\mu^2_N$)} \\
              &  $J_i$ $\rightarrow J_f$  & Exp. & Ref.     & QPM$_{\rm p}$        & QPM$_{\rm n}$ & Exp.  & Ref.   & QPM \\
\hline
              & 3$_1^-$ $\rightarrow$ 0$_1^+$  &  24(3) & \cite{kibedi2002}   &  19.7     & 31.5   &       &          &     \\
$^{94}$Mo     & 3$_2^-$ $\rightarrow$ 0$_1^+$  &   &                   &  2.4      & 3.8           &     &     &     \\
              & 3$_2^-$ $\rightarrow$ 3$_1^-$  & $<$35 & \cite{fransen2003}    &  51.6     &               & 0.39(7) & \cite{fransen2003} & 0.17 \\
              & 5$_1^-$ $\rightarrow$ 0$_1^+$  &  &                    &  6.8      & 6.8           &    &      &     \\
\hline
              & 3$_1^-$ $\rightarrow$ 0$_1^+$  &  19(6) & \cite{kibedi2002}   &      17.2     & 26.7          &    &      &     \\
$^{92}$Zr     & 3$_2^-$ $\rightarrow$ 0$_1^+$  &   &   &  1.4      & 1.7           &     &     &     \\
              & 3$_2^-$ $\rightarrow$ 3$_1^-$  &   &                   &  20.0     &    &           &          &     \\
              & 5$_1^-$ $\rightarrow$ 0$_1^+$  &   &                   &  5.2      & 3.2           &     &     &     \\
\hline
$^{94}$Zr     & 3$_1^-$ $\rightarrow$ 0$_1^+$  &  24(8) & \cite{kibedi2002}    &  20.5      & 38.7    &      &          &     \\
              & 5$_1^-$ $\rightarrow$ 0$_1^+$  &    &                  &   7.5      & 10.2          &      &    &     \\
\hline
\hline
\end{tabular}
\end{threeparttable}
\end{table}

In the QPM, the properties of the 2qp states are independent of the strength of the residual interaction and are solely determined by the parameters of the Wood-Saxon potential and the pairing force. 
These parameters are adjusted to describe the properties of nuclei in this mass region. 
Therefore, it is unlikely that the QPM overestimates the 2qp energies by more than 1 MeV. 
The ($n,n^{\prime}\gamma$) experiments described in Refs.~\cite{fransen2003, fransen2005, elhami2008}, therefore, most likley missed the main fragments of the octupole MSS.

\subsubsection{Proton and electron scattering}
\label{3- and 5- Proton and electron scattering}

\begin{figure}
\centering
    \includegraphics[width=\columnwidth]{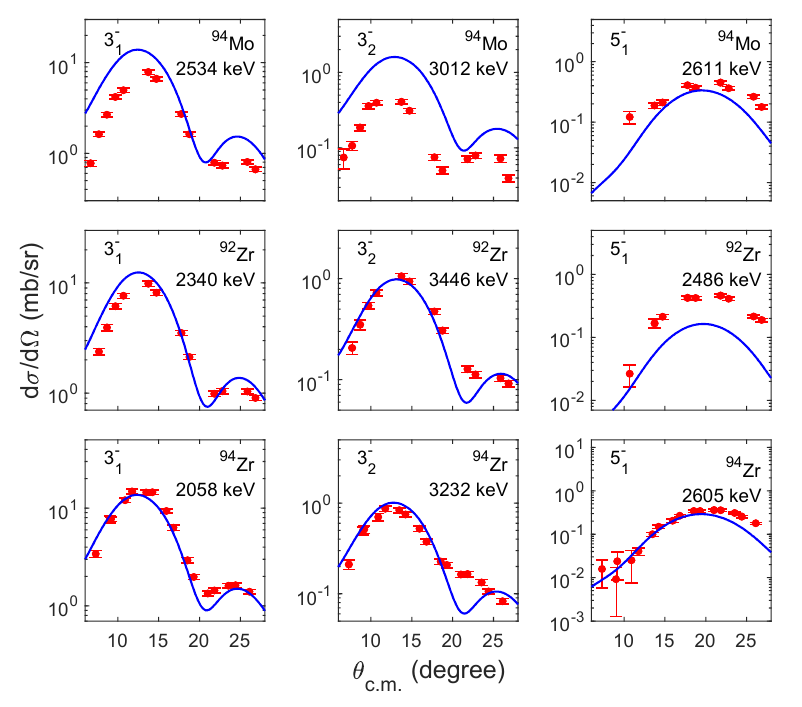}
    \caption[Comparison of the calculated and measured proton scattering cross sections of $3^-$ and $5^-$ states.]
    {\label{Fig: pp comparison of octupole states} Same as Fig.~\ref{Fig: pp comparison of quadrupole states} but for $3^-$ and 5$^-$ states.}
\end{figure}

\begin{figure}
\centering
    \includegraphics[width=0.73\columnwidth]{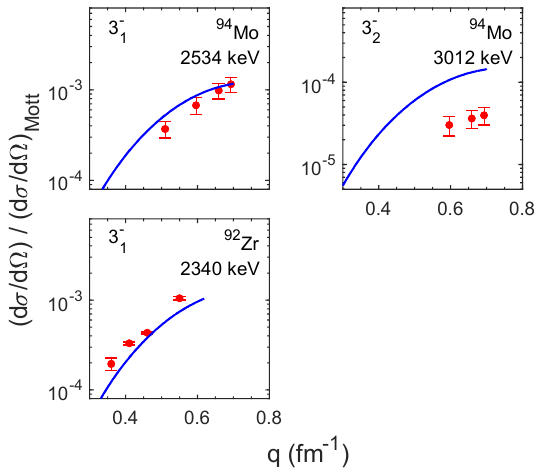}
    \caption[Comparison of the calculated and measured electron scattering cross sections of $3^-$ states.]
    {\label{Fig: ee comparison of octupole states} Same as Fig.~\ref{Fig: ee comparison of quadrupole states} but for octupole states.}
\end{figure}

Figure~\ref{Fig: pp comparison of octupole states} compares the experimental proton scattering cross sections exciting the $3^-_{1,2}$ and $5^-_1$ states to the QPM predictions.
Shape and magnitude of the octupole excitations are well described for $^{92,94}$Zr, while absolute values are somewhat overestimated for $^{94}$Mo.
The angular distributions of the $5^-_1$ states are well accounted for and absolute cross sections are  reasonably reproduced for $^{94}$Mo and $^{94}$Zr, but overestimated for $^{92}$Zr.
The electron scattering data available for the $3^-_1$ states of $^{94}$Mo and $^{92}$Zr shown in Fig.~\ref{Fig: ee comparison of octupole states} can be well described. 
This indicates that the proton part of their wave functions is correct, and the differences for the proton scattering result of $^{94}$Mo originate from the description of the neutron part.

In Ref.~\cite{scheck2010}, a $J=3$ state at 3040 keV is discussed as a candidate for a 3$^-$ MSS of $^{92}$Zr.
No parity is known for this state. 
The proton scattering data presented in Paper I do not provide any evidence of a 3$^-$ state at this energy.
However, it would be very close to an observed 2$^+$ state with an energy of 3057 keV, and the energy resolution would not be good enough to separate such a doublet. 
The angular distribution shown in Fig.~6 of Paper I does not show any hint of a contribution from a 3$^-$ state. 
This points either to a $3^+$ state or to a $3^-$ state with very small g.s.\ excitation strength.
The occurrence of a $3^-$ one-phonon MSS must remain open at this point. 
Most likely, it is considerably fragmented. 

\subsection{Description of $4^+$ states}
\label{subsec: Description of $4^+$ states}

\subsubsection{Energies and transition strengths}
\label{subsubsec: 4+ Energies and transition strengths}

\begin{table}
\centering
\renewcommand*{\arraystretch}{1.3}
\caption[Structure of the hexadecapole QPM wave functions in terms of QRPA-phonons.]{\label{tab: hexadecapole QPM wave functions}
Same as Tab.~\ref{tab: quadrupole QPM wave functions} but for hexadecapole QPM states.
The microscopic structure of the QRPA hexadecupole phonons is provided in Table \ref{tab: hexadecapole QRPA wave functions}.}
\begin{tabular}{cccccc}
\hline
\hline
 &    $J^\pi$            & \multicolumn{2}{c}{$E$ (keV)} & \multicolumn{2}{c}{Structure} \\
             &                         & Exp.          & QPM &  &  \\
\hline
           & 4$_1^+$  & 1574 & 1205  & 43\%$[4^+_1]$ + 44\%$[2^+_1 \otimes 2^+_1 ]$ & \\
$^{94}$Mo  & 4$_2^+$  & 2295 & 1976  & -28\%$[4^+_1]$ + 10\%$[4^+_2]$ - 10\%$[2^+_1 \otimes 2^+_1]$  & \\
          & 4$_4^+$  & 2768 & 2699  & 43\%$[4^+_2]$ - 24\%$[2^+_1 \otimes 2^+_1 ]$ & \\
\hline
        & 4$_1^+$  & 1496 & 1495  & 56\%$[4^+_1]$ + 29\%$[2^+_1 \otimes 2^+_1 ]$   & \\
$^{92}$Zr    & 4$_2^+$  & 2398 & 1951  & -36\%$[4^+_1]$ + 6\%$[4^+_2]$ + 24\%$[2^+_1\otimes 2^+_1 ]$    & \\
    & 4$_3^+$  & 2865 & 2643  & 37\%$[4^+_2]$ + 34\%$[2^+_1 \otimes 4^+_1 ]$    & \\
\hline
$^{94}$Zr  & 4$_1^+$  & 1470 & 1404  &  76\%$[4^+_1]$ +16\%$[2^+_1 \otimes 2^+_1]$ &\\
    & 4$_2^+$  & 2330 & 2584  &  48\%$[4^+_2]$ +17\%$[2^+_1 \otimes 2^+_2]$ & \\
\hline
\hline
\end{tabular}
\end{table}

Table~\ref{tab: hexadecapole QPM wave functions} presents the QPM wave functions of the lowest 4$^+$ states of $^{92,94}$Zr and $^{94}$Mo. 
The leading components are the [4$^+_1]$, [4$^+_2]$ and $[2^+_1 \otimes 2^+_1 ]$ similar to the quadrupole states discussed in Sec.~\ref{subsec: 2+ states}, but the three important phonons are much closer in energy leading to a strong mutual mixing. 
For example, the [4$^+_1]$ and $[2^+_1 \otimes 2^+_1 ]$ phonons contribute with comparable amplitudes to the 4$_1^+$ state of $^{94}$Mo. 
This is in contrast to the wave functions of the 2$^+$ states, where each of the three phonons dominates one QPM state. 
This complicates the unambiguous identification of fragments of MSS and multiphonon candidates.

The QPM predictions of the $4^+$ state energies shown in Tab.~\ref{tab: hexadecapole QPM wave functions} show deviations of a few hundred keV for some states. 
Especially in $^{94}$Zr, the QPM predicts the 4$^+_2$ state to lie at an energy of $\sim$1500 keV (not shown in Tab.~\ref{tab: hexadecapole QPM wave functions}), i.e. very close to the 4$^+_1$ state in strong disagreement with experiment. 
Hence, the properties of the third 4$^+$ state of the QPM are compared to the experimental 4$^+_2$ state in Tab.~\ref{tab: hexadecapole transition matrix elements} and Fig.~\ref{Fig: pp comparison of hexadecapole states}. 
Due to these problems, $^{94}$Zr is excluded from the following discussion. 
With the unperturbed energies of the three dominant phonons close to each other, the final QPM wave functions are highly sensitive to details of the calculations. 
Already, small variations of the strength of the residual interaction change the phonon structure of the final QPM wave functions and the corresponding energies and transition strengths significantly. 

\begin{table} 
\centering
\caption[Comparison of the calculated and measured transition strengths of hexadecapole states.]
{\label{tab: hexadecapole transition matrix elements} Same as Tab.~\ref{tab: quadrupole transition matrix elements} but for hexadecapole states. 
Experimental data are from Refs.~\cite{fransen2003, fransen2005, elhami2008, elhami2013}.  
QPM $M1$ transitions strengths between one-phonon 4$^+$ configurations only are shown in brackets marked with an asterisk.}
\begin{threeparttable}[c]
\begin{tabular}{ccccccc}
\hline
\hline
              &                                & \multicolumn{3}{c}{$B(EJ)$(W.u.)} & \multicolumn{2}{c}{$B(M1)$($\mu^2_N$)} \\
$^{\rm A}X$   & $J_i\rightarrow J_f$      & Exp.                  & QPM$_{\rm p}$ & QPM$_{\rm n}$ & Exp.      & QPM   \\
\hline
      & 4$_1^+$ $\rightarrow$ 0$_1^+$  &                      &  2.1          & 16.2          &    & \\
              & 4$_2^+$ $\rightarrow$ 0$_1^+$  &                      &  0.81         &  1.1          &          &     \\
              & 4$_4^+$ $\rightarrow$ 0$_1^+$  &                      &  1.93         &  5.8          &          &     \\
$^{94}$Mo     & 4$_2^+$ $\rightarrow$ 4$_1^+$  & 1.2(33)              &  1.7          &           & 1.23(30) & 0.30 (0.12$^*$) \\
              & 4$_3^+$ $\rightarrow$ 4$_1^+$  &                      &  1.6          &           & 0.23(6)  & 0.82 (0.62$^*$) \\
              & 4$_4^+$ $\rightarrow$ 4$_1^+$  & 36$^{+5}_{-4}$       &  0.3          &           & 0.090(11)& 0.23 (0.63$^*$) \\
              & 4$_1^+$ $\rightarrow$ 2$_1^+$  & 26.0$^{+4.2}_{-3.2}$ &  12.6         &               &          &     \\
              & 4$_2^+$ $\rightarrow$ 2$_1^+$  & 5.9$^{+1.0}_{-0.8}$  &  2.7          &               &          &     \\
\hline
              & 4$_1^+$ $\rightarrow$ 0$_1^+$  &                      & 1.4           &   5.0         &          &     \\
              & 4$_2^+$ $\rightarrow$ 0$_1^+$  &                      & 0.15          &   0.53        &          &     \\
              & 4$_3^+$ $\rightarrow$ 0$_1^+$  &                      & 1.4           &   1.9         &          &     \\
$^{92}$Zr     & 4$_2^+$ $\rightarrow$ 4$_1^+$  & 2.3(11)              & 0.2           &               & 0.26(3)  & 0.32\\
              & 4$_1^+$ $\rightarrow$ 2$_1^+$  & 4.05(11)             & 3.8           &               &          &     \\
              & 4$_2^+$ $\rightarrow$ 2$_1^+$  & 6.1(8)               & 3.1           &               &          &     \\
              & 4$_3^+$ $\rightarrow$ 2$_1^+$  &                      & 0.1           &               &          &     \\
              & 4$_4^+$ $\rightarrow$ 2$_1^+$  &                      & 2.0           &               &          &     \\
\hline
              & 4$_1^+$ $\rightarrow$ 0$_1^+$  &                      &   1.1         &    8.4        &          &     \\
              & 4$_2^+$ $\rightarrow$ 0$_1^+$  &                      &   1.9         &    6.0        &          &     \\
$^{94}$Zr     & 4$_2^+$ $\rightarrow$ 4$_1^+$  &                      &               &               &          &0.2  \\
              & 4$_1^+$ $\rightarrow$ 2$_1^+$  & 0.880(23)            &   0.9         &               &          &     \\
              & 4$_2^+$ $\rightarrow$ 2$_1^+$  & 13$^{+4}_{-7}$       &   3.3         &               &          &     \\
\hline
\hline
\end{tabular}
\end{threeparttable}
\end{table}

The experimental and theoretical transition strengths are shown in Tab.~\ref{tab: hexadecapole transition matrix elements}. 
Additionally, the $g$ factor of the $4^+_1$ state of $^{92}$Zr has been measured \cite{benczerkoller2007}. 
Its large negative value is well reproduced by the QPM (cf.~Tab.~\ref{tab: hexadecapole g-factors}) indicating a dominant neutron component in the wave function.
The 4$^+_1$ and 4$^+_2$ states of $^{94}$Mo and $^{92}$Zr are experimentally connected by large $B(M1)$ values of 1.23(30)~$\mu_N^2$ and
0.26(3)~$\mu_N^2$, respectively. 
Also, the $B(M1;4^+_3\rightarrow4^+_1)=0.23(6)\mu_N^2$ transition in $^{94}$Mo is sizable. 
To answer whether the 4$^+_2$ and 4$^+_3$  states in $^{94}$Mo and the 4$^+_2$ state in $^{92}$Zr carry fractions of a one-phonon hexadecapole MSS, it is useful to look at the diagonal $M1$ transition rates between the dominantly involved pure one-phonon or two-phonon configurations. 
For $^{94}$Mo, one finds $B(M1;[4^+_1]\rightarrow[4^+_1])=0.35\mu_N^2$, 
$B(M1;[4^+_2]\rightarrow[4^+_1])=1.2\mu_N^2$, and 
$B(M1;[2^+_1 \otimes 2^+_1 ]\rightarrow[2^+_1\otimes 2^+_1 ])=0.9\mu_N^2$. 
They all have similar size. 
Thus, in the case of strong phonon mixing predicted by the QPM for the three nuclei considered, large $M1$ transition strengths between them do not enable an unambiguous discrimination between the contribution from the one-phonon hexadecupole MSS or from the splitting of FS two-phonon configurations over neighboring eigenstates. 
There are indications in the experimental data supporting strong phonon mixing. 
The 4$^+_1$ state of $^{94}$Mo decays with  large $B(E2)$ value of 26 W.u.\ to the 2$^+_1$ state indicating a significant $[2^+_1\otimes 2^+_1 ]$ component. 
Also, the 4$^+_2$ state has a sizable $B(E2)$ value of about 6 W.u.\ to the 2$^+_1$ state most likely caused by a $[2^+_1\otimes 2^+_1 ]$ component in the wave function. 
Similar observations can be made for the 4$^+_1$ and 4$^+_2$ states of $^{92}$Zr.

\begin{table} 
\centering
\caption[Comparison of the calculated and measured $g$-factors of hexadecapole states.]
{\label{tab: hexadecapole g-factors} Same as Tab.~\ref{tab: quadrupole g-factors} but for hexadecapole states.}
\begin{tabular}{cccc}
\hline
\hline
$^{\rm A}X$   &  $g(J^{\pi})(\mu_n$)      & Exp.       & QPM     \\
\hline
$^{92}$Zr     &  $g(4^+_1$)                 & -0.5(1)   & -0.55      \\
\hline
\hline
\end{tabular}
\end{table}

The QPM approach has problems to describe the experimental transition strengths in Tab.~\ref{tab: hexadecapole transition matrix elements}. 
The important $B(M1;4^+_2\rightarrow4^+_1$) value is underestimated by a factor of four in $^{94}$Mo. 
Shown in brackets and marked with an asterisk is the part stemming from the one-phonon components only, i.e. more than 50$\%$ of the $B(M1;4^+_2\rightarrow4^+_1$) value is predicted to be due to transitions between two-phonon components. 
Furthermore, the $B(E2;4^+_1\rightarrow2^+_1$)  and $B(E2;4^+_2\rightarrow2^+_1$) values are both underestimated by more than a factor of two. 
This is a clear indication that the QPM amplitude of the $[2^+_1\otimes2^+_1 ]$ component in both states is too small.
An increase of this component would also improve the description of the $B(M1:4^+_2\rightarrow4^+_1$) value since the diagonal $M1$ matrix element of the $[2^+_1\otimes 2^+_1 ]$ phonon is large. 
On the other hand, the excellent description of the proton and electron scattering cross sections of the 4$^+_1$ state discussed below would become worse. 
A simple redistribution of the phonons does not seem to improve the overall description of the experimental observables.
This points to deficiencies describing the internal structure of the hexadecapole phonons in $^{94}$Mo.
A possible identification of the main components of the hexadecapole MSS would require a model that describes all important experimental observables sufficiently well and thereby provides the structure of the wave functions. 

The QPM is more successful in describing the properties of the 4$^+$ states of $^{92}$Zr.
The $B(E2;4^+_1\rightarrow2^+_1$)  as well as the $B(M1;4^+_2\rightarrow4^+_1)$ transition strength are in good agreement with the data. 
The measured $B(E2;4^+_2\rightarrow2^+_1$) value, however, is again a factor of two larger than the QPM result. 
Interestingly, in $^{92}$Zr, the main component of the $[2^+_1\otimes 2^+_1 ]$ phonon is found in the second rather than the first 4$^+$ state as in $^{94}$Mo. 

\subsubsection{Proton and electron scattering}
\label{subsubsec: 4+ Proton and electron scattering}

\begin{figure}
\centering
    \includegraphics[width=\columnwidth]{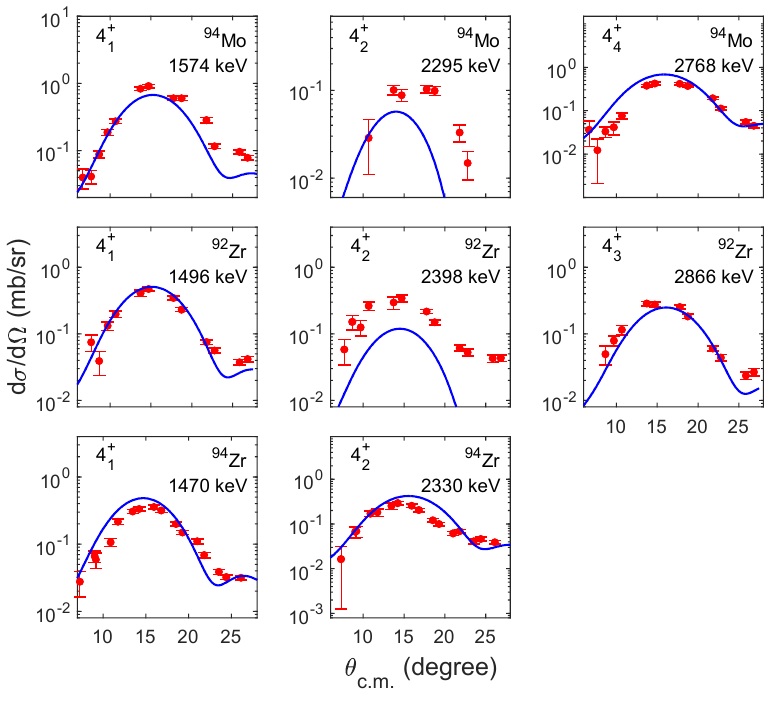}
    \caption[Comparison of the calculated and measured proton scattering cross sections of hexadecapole states.]
    {\label{Fig: pp comparison of hexadecapole states} Same as Fig.~\ref{Fig: pp comparison of quadrupole states} but for hexadecapole states.}
\end{figure}

\begin{figure}[b]
\centering
    \includegraphics[width=\columnwidth]{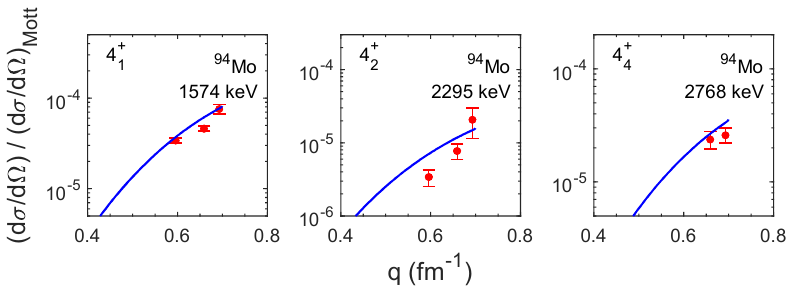}
    \caption[Comparison of the calculated and measured electron scattering cross sections of hexadecapole states.]
 {\label{Fig: ee comparison of hexadecapole states} Same as Fig.~\ref{Fig: ee comparison of quadrupole states} but for hexadecapole 
states.}
\end{figure}

The proton and electron scattering cross sections shown in Figs.~\ref{Fig: pp comparison of hexadecapole states} and \ref{Fig: ee comparison of hexadecapole states} are evidence of large one-phonon components in the wave function. 
In both reactions, cross sections are large and the proton scattering cross sections are described well considering a one-step excitation mechanism only. 
One exception is the theoretical angular distribution of the 4$^+_2$ state of $^{94}$Mo, which strongly disagrees with the experimental data. 
In contrast to all other 4$^+$ states, the first maximum of the data is shifted to larger scattering angles. 
The theoretical curve is influenced by a destructive interference of the [4$^+_1]$  and [4$^+_2]$ phonons both contributing with sizable amplitudes to the 4$^+_2$ state.
The sensitivity of this effect to details of the calculations has been discussed above.

The QPM predicts the main fragment of the mixed-symmetric hexadecupole phonon in the 4$^+_4$ state of $^{94}$Mo (Table~\ref{tab: hexadecapole QPM wave functions}). 
The description of the corresponding electron and proton scattering cross sections is fair but, as discussed above, the $M1$ part of the transition to the $4^+_1$ state is overestimated, while the large $E2$ transition strength is underestimated by two orders of magnitude. 
Although the proton scattering cross sections of the 4$^+_1$ and 4$^+_3$ states in $^{92}$Zr are very well described in terms of shape and magnitude, the cross section of the 4$^+_2$ state is significantly underestimated which hints at a larger one-phonon component to its wave function than estimated by the QPM.

%Its value of -0.5(1) is larger than the g-factor of the 2$^+_1$ state of -0.18(1). This supports the picture given at the beginning of the section that the two 2qp-states (2d$_{5/2} \otimes$ 2d$_{5/2})_n$ and (1g$_{9/2} \otimes$ 1g$_{9/2})_p$ are weaker admixed in case of the 4$^+$ states than for the 2$^+$ states due to the different               
%strengths of the quadrupole and hexadecapole residual interactions.  

\subsubsection{Two-state model}
\label{subsubsec: 4+ two-state model}

Table~\ref{tab: hexadecapole QRPA wave functions} shows the structure of the first and second hexadecapole QRPA phonons of $^{92,94}$Zr and $^{94}$Mo. 
The same  ($2d_{5/2}\otimes 2d_{5/2})_n$ and ($1g_{9/2}\o\times 1g_{9/2})_p$ 2qp states as for the quadrupole states are important. 
Again, one can understand the structure of the phonons in the simple two-state model introduced for the 2$^+$ states. 

\begin{table}
\caption[Structure of the hexadecapole QRPA-phonons in terms of two-quasiparticle states.]{\label{tab: hexadecapole QRPA wave functions} Same as Tab.~\ref{tab: quadrupole QRPA wave functions} but for hexadecapole QRPA-phonons.}
\begin{tabular}{cccccccc}
\hline
\hline
             & & \multicolumn{3}{c}{($2d_{5/2}\otimes 2d_{5/2})_n$}  & \multicolumn{3}{c}{($1g_{9/2}\otimes 1g_{9/2})_p$}  \\
$^{\rm A}X$  & QRPA phonons        &$\psi$ &$\phi$ & $\%$  & $\psi$   & $\phi$  & $\%$  \\
\hline
$^{94}$Mo    & $[4^+_1]$  & 1.34      & 0.04    & 90.30   & 0.41      & 0.08        & 8.17   \\
             & $[4^+_2]$  & -0.44     & 0.05    & 9.43    & 1.30      & 0.06        & 84.95  \\
\hline
$^{92}$Zr    & $[4^+_1]$  & 1.39      & 0.04    & 96.22   & 0.22      & 0.06        & 2.21  \\
             & $[4^+_2]$  & -0.26     & 0.05    & 3.32    & 1.29      & 0.06        & 83.47 \\
\hline
$^{94}$Zr    & $[4^+_1]$  & 1.40      & 0.04    & 98.14   & 0.15      & 0.05        & 0.93  \\
             & $[4^+_2]$  & -0.18     & 0.05    &  1.51   & 1.00      & 0.06        & 50.19 \\
\hline
\hline
\end{tabular}
\end{table}

Three important differences are apparent in comparison to the quadrupole case shown in Tab.~\ref{tab: quadrupole QRPA wave functions}. 
First, the mixing between both 2qp states is much weaker. 
The [4$^+_1]$ phonons of all three nuclei are almost pure ($2d_{5/2}\otimes 2d_{5/2})_n$ states with small proton admixtures. 
This can be explained by much weaker hexadecapole proton-neutron residual interactions. 
Like the 2$^+$ case, both dominant 2qp configurations show stronger mixing in the case of $^{94}$Mo than in $^{92,94}$Zr due to the reduced energy difference between the proton and neutron 2qp states.
Second, other 2qp states play a minor role only. 
Their contribution to the norm of the wave functions is less than 15$\%$ for each phonon (except for the [4$^+_2]_{[RPA]}$ phonon of $^{94}$Zr), i.e., the simple two-state model is a good approximation of the QRPA wave functions.
Furthermore, many 2qp states close in energy to the two dominant 2qp configurations cannot couple to angular momentum $J=4$.
Third, the [2$^+_1$$\otimes$2$^+_1$)] two-phonon state is found at comparable excitation energy.
In the collective picture, the [4$^+_1$] state is a pure two-phonon state.

\subsection{Discussion}
\label{subsec: Discussion}

The ability of the QPM to describe natural-parity states in the $Z=40$,$N=50$ region was tested by an extensive comparison to experimental data in $^{92,94}$Zr and $^{94}$Mo. 
The available observables include g.s.\ properties, excitation energies, electromagnetic moments and transition strengths, as well as proton and electron scattering cross sections.

The description of the 2$^+$ states of $^{92}$Zr and $^{94}$Mo is excellent. 
Most of the observables are described with reasonable accuracy providing confidence in the QPM wave functions of those states. 
The most notable discrepancy is an underestimation of the charge transition radii of the 2$^+_1$ and 2$^+_3$ states of $^{94}$Mo. 
Also, the predicted energy of the 2$^+_2$ state of $^{94}$Mo is off by $\sim 400$ keV. 
Because of the restriction to the calculations on the RPA level, it was necessary to lower the energy of the
($2d_{5/2}\otimes 2d_{5/2})_n$ 2qp state in $^{94}$Zr artificially by 200 keV in order to achieve a reasonable $B(E2)$ value of the quadrupole MSS. 
A better description of $^{94}$Zr would require the modification of the parameters of the Woods-Saxon potential. 
Such empirical corrections of the single-particle spectrum are quite common, see e.g.\ the shell-model study of MSS for nuclei in the vicinity of the $N = 82$ shell closure \cite{sieja2009}. 
In general, the proton and neutron 2qp states of $^{94}$Zr seem to be more separated in energy than in $^{92}$Zr and $^{94}$Mo.

%\begin{figure*}
%\centering
%    \includegraphics[width=\textwidth]{schematic_structure.pdf}
%    \caption[Unperturbed phonon- and two-quasiparticle-energies in $^{94}$Mo.]
%    {\label{Fig: schematic structure} Unperturbed phonon (solid blue lines) and 2qp-energies (black dashed lines)
%     of 2$^+$, 3$^-$ and 4$^+$ states in $^{94}$Mo. } % (add 3-2+1)                
%\end{figure*}

For 3$^-$ and 5$^-$ states, less experimental data are available. 
They are reasonably well reproduced by the calculations. 
The QPM results suggest that an experimental identification of the octupole one-phonon MSS is difficult. 
Due to the absence of suitable 2qp states at low energies, the mixed-symmetric QRPA phonon is predicted at energies $>4$ MeV in all three nuclei. 

The calculations also predict a hexadecapole MSS. 
The formation mechanism is similar to the quadrupole case although the mixing of the two important 2qp states is much weaker, and the occurrence of the $2^+ \times 2^+$ two-quadrupole phonon $4^+$ state close to, or even below, the energies of the leading 2qp configurations further complicates the analysis.
While some features of the data are well described, the QPM calculations are not able to achieve a good description of all observables simultaneously.
Because of the sensitivity of the theoretical results to details of the model discussed in Sec.~\ref{subsubsec: 4+ Energies and transition strengths}, testing claims of a hexadecapole MSS based on the reproduction of certain experimental observables is not possible.

\section{Signatures of 2$^+$ mixed-symmetry states}
\label{sec: Signatures of 2$^+$ mixed-symmetry states}

In this Section, we discuss possible signatures of MSS other than strong $M1$ transitions between MSS and FSS of the same spin and parity. 
In the first part, the capability of the quantity $R$ defined in Eq.~(\ref{eq: ratio}) to identify a MSS if, a priori, nothing is known about its excitation energy. Secondly, the sensitivity of proton and neutron transition radii to the mixed-symmetry character is discussed.

\subsection{Proton-neutron matrix elements}
\label{subsec: Proton-neutron matrix elements}

The authors of Refs.~\cite{deleo1998,deleo1989,pignanelli1988} used absolute proton and neutron transition matrix elements to evaluate the quantity $R$
\begin{equation}
\label{eq: ratio}
R = \left|\frac{M_n - M_p}{M_n + M_p} \right|.
\end{equation}
Here, $M_p$ and $M_n$ are the proton and neutron transition matrix elements to the ground state. 
They looked for enhanced values of $R$ and identified the corresponding states as MSS. 
For the nuclei $^{92,94}$Zr and $^{94}$Mo, proton and neutron $B(E2)$ values are available. 
In these cases, the MSS candidates are suitably identified from the $B(M1)$ values, and the capability of $R$ to identify members of this class of states can be tested.

\begin{table}
\centering
\caption[QRPA results for the $R$-value using the full QPM model space and only the main components.]
{\label{tab: ratio}
QRPA results for the $R$-value defined in Eq.~\ref{eq: ratio} using only the two main components (`Valence') and the full QPM model space (`Full').}
\begin{tabular}{ccccccc}
\hline
\hline
             & \multicolumn{2}{c}{$^{94}$Mo}  & \multicolumn{2}{c}{$^{92}$Zr} & \multicolumn{2}{c}{$^{94}$Zr}    \\
             &  2$^+_1$ & 2$^+_{\rm ms}$ & 2$^+_1$ & 2$^+_{\rm ms}$ & 2$^+_1$ & 2$^+_{\rm ms}$   \\
\hline
Valence      & 0.03     & 6.16           & 0.39    & 14.67          & 0.37    & 4.72            \\
Full         & 0.16     & 0.02           & 0.21    & 0.08           & 0.27    & 0.14             \\
\hline
\hline
\end{tabular}
\end{table}

In the two-state model introduced in Sec.~\ref{subsubsec: Two-state model}, one expects $R \ll 1$ for the 2$^+_1$ state due to the cancellation of proton and neutron matrix elements in the numerator. 
For the MSS, both matrix elements add coherently in the nominator and cancel in the numerator leading to $R>1$. 
The corresponding QPM results - considering the two main neutron and proton components only - are shown in Tab.~\ref{tab: ratio} in the row labeled `Valence'. Clearly, the MSS $R$ value is enhanced in comparison to the 2$_1^+$ state. 
Moreover, it is always $\gg 1$ possibly allowing these states to be differentiated from other non-collective states of the same multipolarity. 
For noncollective states, one can expect a dominant proton or neutron 2qp component leading to $R\approx$1. 
However, the applicability of $R$ to identify MSS depends on the contributions of other 2qp states not considered in the two-state model, which can modify the final value of $R$ substantially.

Due to the good agreement in the case of the quadrupole states discussed in the previous section, the following analysis is based on the QPM results.
The corresponding $R$ values are displayed in Tab.~\ref{tab: ratio} in the row labeled `Full'. 
The $R$ values of the 2$^+_1$ states fulfill expectations. 
Surprisingly, the $R$ values of the MSS states are not only smaller than one, they are even smaller than those of the 2$^+_1$ states.

\begin{figure}
\centering
    \includegraphics[width=0.95\columnwidth]{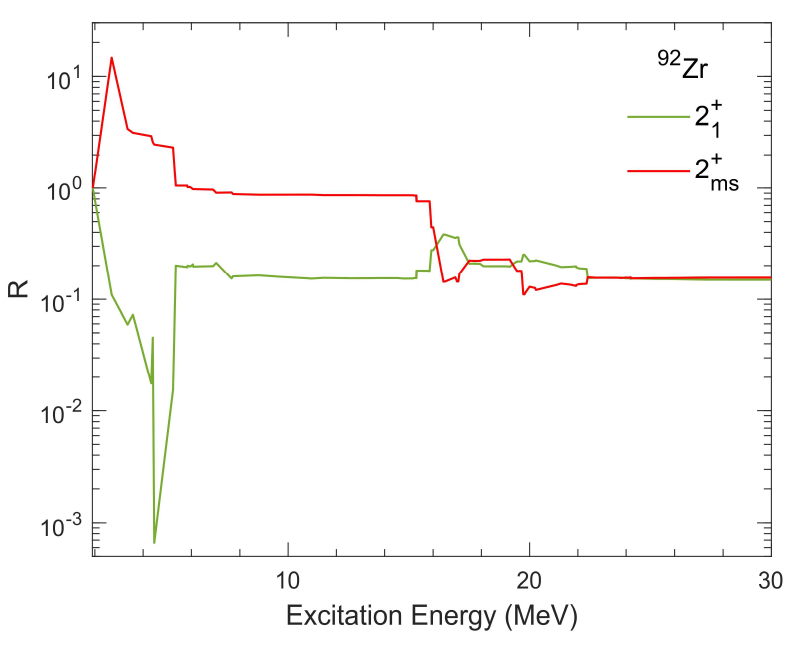}
    \caption[Running sum of the $R$ value of $^{92}$Zr.]
    {\label{Fig: ratio} Running sums of the $R$ value defined in Eq.~(\ref{eq: ratio}) for the [2$^+_1]$ and [2$^+_{\rm ms}]$ phonons in $^{92}$Zr.}
\end{figure}
In order to understand this unexpected behavior, $R(2^+_1$) and $R(2^+_{\rm ms}$) for $^{92}$Zr are decomposed in Fig.~\ref{Fig: ratio} into the contributions of the various 
2qp components according to their RPA excitation energy.
All 2qp components up to a given excitation energy are included in the calculation of $R$ while those above are neglected. 
The value of $R(2^+_1$) drops quickly to $R \approx 0$. The large value of $R(2^+_{\rm ms}$) obtained when only the two main components are considered is already significantly reduced by other 2qp states at energies $<5$ MeV. 
Still, one finds values $R > 1$.
A restriction to 2qp states in the valence shell, where one has to introduce effective charges, is the domain of the shell model. 
States between 5 and 15 MeV contribute little and the $R$ values stay rather constant with $R(2^+_{\rm ms})\approx$1.
Between 16 and 22 MeV, it drops eventually to $R \ll 1$ due to strong mixing with states of the isoscalar giant quadrupole resonance (ISGQR).
The underlying mechanism is explained in Sec.~\ref{sec:IV}.
Obviously, $R$ cannot be considered as a good signature to identify MSS, and the conclusions of Refs.~\cite{deleo1989,deleo1998,pignanelli1988} are questioned.

Nevertheless, it might be worthwhile to test the wave function of a known MSS by comparison of its proton and neutron $B(E2)$ values.
For example, the total neutron transition matrix element of the mixed-symmetric QRPA phonon of $^{94}$Mo amounts to 5.93~efm$^2$, while the contribution of the ($2d_{5/2}\otimes2d_{5/2})_n$-2qp-state is -3.43~efm$^2$;
i.e., the main component still has a pronounced influence on the total neutron $B(E2)$ value. 
Consequently, a successful description of absolute proton and electron scattering cross sections as discussed in Secs.~\ref{subsubsec: 2+ proton scattering} and \ref{subsubsec: 2+ Electron scattering} is also an indication that the model predicts the properties of the main MSS components correctly.

\subsection{Proton-neutron transition radii}
\label{subsec: Proton-neutron transition radii}

In this section, we discuss how the differences of quadrupole proton and neutron transition radii can provide a signature of $2^+$ MSS independent of absolute electromagnetic transition strengths.
The idea has been briefly discussed for $^{94}$Mo \cite{burda2007} and $^{92}$Zr \cite{walz2011}.
Here, we provide an in-depth analysis and extend it to $^{94}$Zr.
Furthermore, this new observable 
permits a direct determination of the relative sign between the main proton and neutron 2qp configurations in the wave function of the MSS.

\begin{figure}
\centering
    \includegraphics[width=0.9\columnwidth]{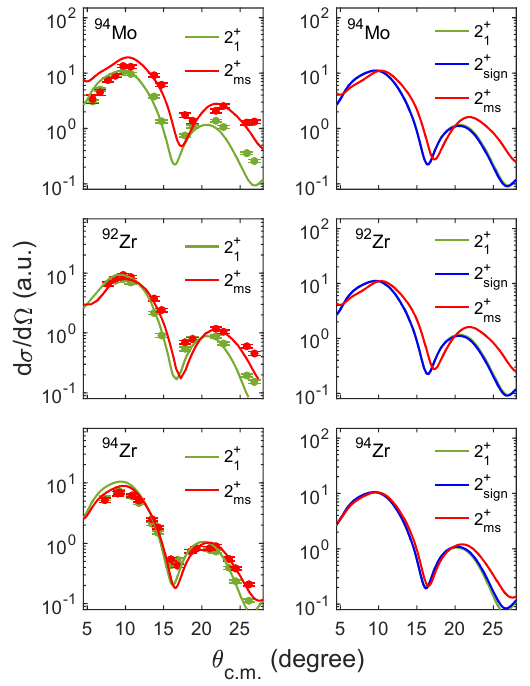}
    \caption[Cross sections of the 2$^+_1$ and 2$^+_{\rm ms}$ states in $^{92,94}
    $Zr and $^{94}$Mo.]
    {\label{Fig: 2+_shift_comparison} Cross sections of the 2$^+_1$ and 2$^+_{\rm\
    ms}$ states in $^{92,94}$Zr and $^{94}$Mo. The cross sections of the 
    mixed-symmetry states are scaled to the one of the corresponding 2$^+_1$ states 
    in order to allow a comparison of their shapes.}
\end{figure}

The left side of Fig.~\ref{Fig: 2+_shift_comparison} compares the measured proton scattering cross sections of the FSS and the MSS $2^+$ states in $^{92,94}$Zr and $^{94}$Mo.
Their angular (or momentum transfer) dependence is closely related to the transition radii~\cite{burda2007}, i.e., they provide information on the radial structure of the excitation in contrast to transition matrix elements. 
The comparison reveals that both have similar shapes but that the maxima and minima of the 2$^+_{\rm ms}$ states are shifted to higher scattering angles (or momentum transfers) with respect to those of the 2$^+_1$ states.
For all three investigated nuclei, a smaller transition radius of the 2$^+_{\rm ms}$ state than that of the 2$^+_1$ state is indicated.
The QPM calculations can quantitatively reproduce this behavior.

In general, proton and neutron transition densities have different transition radii. 
Proton scattering is sensitive to an isoscalar (or matter) transition radius depending on the properties of both densities. 
However, the neutron contribution is expected to dominate for the following reasons. 
First, the effective proton-nucleus interaction at 200 MeV is slightly stronger for neutrons than for protons \cite{love1981,franey1985}. 
Second, the QPM transition densities of the 2$^+$ states are consistent with a small neutron skin.
Since proton scattering at the momentum transfers discussed here mainly probes the surface of the nucleus, the neutron contribution is larger.

The dependence of proton and neutron transition densities on the sign between the leading 2qp components is illustrated in the r.h.s.\ of Fig.~\ref{Fig: 2+_shift_comparison}. 
Here, the QPM results are normalized to each other at the first maximum of the angular distribution.
The blue curves show the QPM results for the MSS with the relative sign between the leading 2qp proton and neutron configurations (cf.\ Tab.~\ref{tab: quadrupole QRPA wave functions} artificially changed from minus to plus. 
They are super-imposable with the $2^+_1$ results making the blue and green curves indistinguishable.

Further insights can be obtained from a theoretical analysis of the transition radii $R_{\rm tr}$. 
For $E2$ transitions,
\begin{equation}
    \label{Eq: transition radius}
    R_{\rm tr} = \sqrt{\frac{\int \rho_{\rm tr} r^4  dr}{\int \rho_{\rm tr}\
    r^2 dr}}.
\end{equation}
Here, $\rho_{\rm tr}$ is the transition density. 
The calculated proton, neutron and 
matter  (defined as the sum) transition radii are summarized in Tab.~\ref{tab: radii}.  
The proton values of the $2^+_1$ and $2^+_{\rm ms}$ states are almost identical, while the neutron values are larger by 0.15 fm for $^{94}$Mo and 0.06 fm for $^{92}$Zr. 
In contrast, for $^{94}$Zr, it is 0.06~fm smaller.
As discussed below, this is partly due to the contribution of the ($2d_{5/2}\otimes3s_{1/2})_n$ 2qp state with an amplitude of 37$\%$.

\begin{table}[b] 
\centering
\caption[Proton, neutron, and matter transition radii of the 2$_1^+$ and 2$_{\rm ms}^+$ states.]{\label{tab: radii} Proton, neutron and matter transition radii of 
the 2$_1^+$ and 2$_{\rm ms}^+$ states. $\Delta R_{\rm m}$ denotes the differences between both matter transition radii of each nucleus. This difference increases from $^{94}$Zr to $^{94}$Mo. }
\begin{tabular}{ccccccccc}
\hline
\hline
             & \multicolumn{2}{c}{$R_{\rm p}$ (fm)} & \multicolumn{2}{c}{$R_{\rm n}$ (fm)} & \multicolumn{2}{c}{$R_{\rm m}$ (fm)} & $\Delta R_{\rm m}\
$ (fm)  \\
             &  2$^+_1$ & 2$^+_{\rm ms}$ & 2$^+_1$ & 2$^+_{\rm ms}$ & 2$^+_1$ & 2$^+_{\rm ms}$   \\
\hline
$^{94}$Mo    &  5.40  &  5.34  &  6.22  &  6.07  &  5.89  &  5.66  &  0.23    \\
$^{92}$Zr    &  5.40  &  5.37  &  6.24  &  6.18  &  5.92  &  5.80  &  0.12    \\
$^{94}$Zr    &  5.40  &  5.38  &  6.31  &  6.37  &  5.99  &  5.97  &  0.02    \\
\hline
\hline
\end{tabular}
\end{table}

\begin{figure} 
\centering
    \includegraphics[width=\columnwidth]{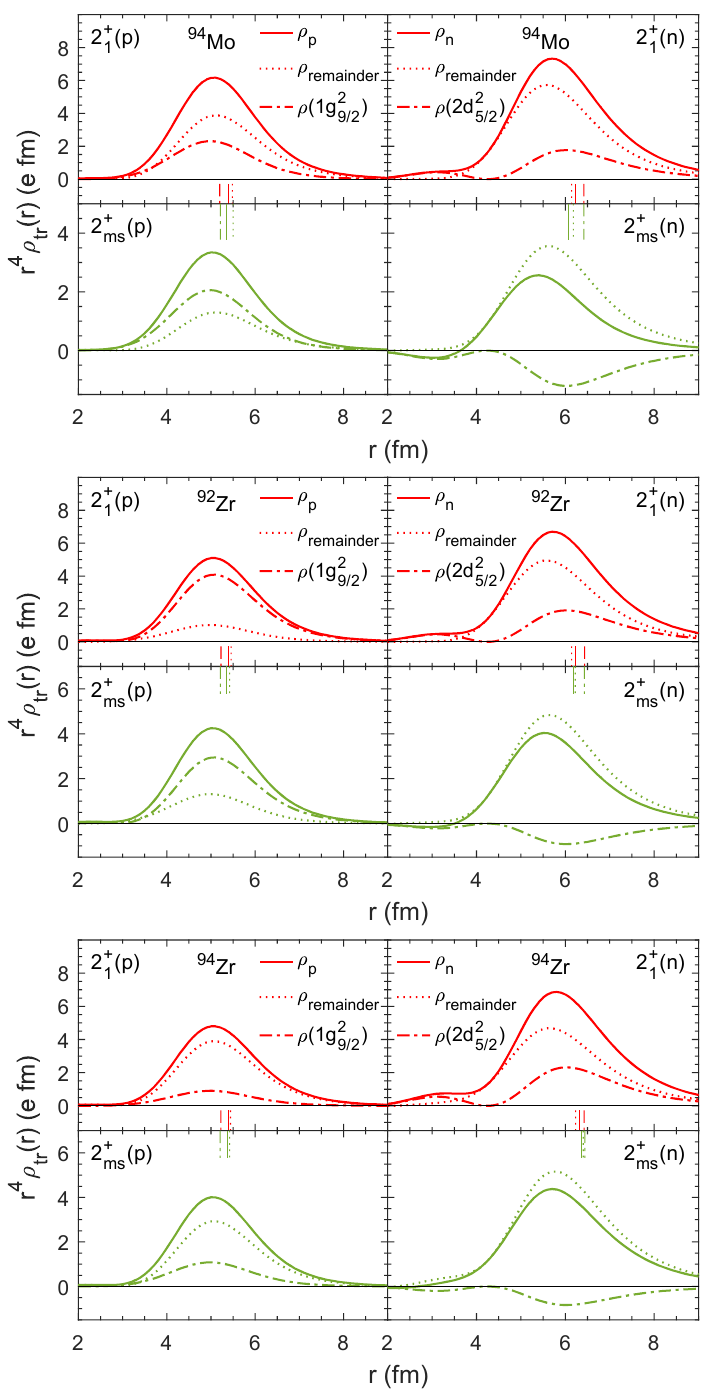}
    \caption[Proton and neutron transition densities weighted with $r^4$ of
     $^{94}$Mo, $^{92}$Zr and $^{94}$Zr.]
{\label{Fig: transdens_shift} Proton and neutron transition densities of the $2^+_1$ (solid red lines) and $2^+$ MSS (solid green lines) states in $^{94}$Mo (top), $^{92}$Zr (middle) and $^{94}$Zr (bottom) weighted with r$^4$. 
Each transition density is decomposed into parts stemming from the main 2qp components (dashed-dotted lines)  and all other 2qp states ($\rho_{\rm remainder}$, dotted lines). 
The vertical lines mark the $R_{\rm tr}$ values of the corresponding densities.}
\end{figure}

The corresponding quantities studied by proton scattering  are the matter transition radii. 
Their difference $\Delta R_{\rm m}$ decreases from $^{94}$Mo to $^{94}$Zr. 
In order to correlate $\Delta R_{\rm m}$ with the change of sign between the main proton and neutron 2qp contributions one has to investigate the theoretical transition densities in detail. 
Figure~\ref{Fig: transdens_shift}
displays the $r^4$-weighted  proton and neutron transition densities of $^{94}$Mo, $^{92}$Zr and $^{94}$Zr for the $2^+_1$ (red lines) and $2^+$ MSS (green lines) states. 
The r$^4$ weighting factor emphasizes the surface region of the nucleus probed by proton scattering. 
The full densities $\rho_{n(p)}$ (solid lines) are decomposed into the contributions of the main ($2d_{5/2}\otimes2d_{5/2})_n$ and ($1g_{9/2}\otimes1g_{9/2})_p$ components (dashed-dotted lines) and the contributions of all other 2qp states $\rho_{\rm remainder}$ (dotted lines). 
The corresponding $R_{\rm tr}$ values of each density are marked by vertical lines.

The `remainder' parts of the proton and neutron densities of the 2$^+_1$ and MSS show similar transition radii. 
The ($1g_{9/2}\otimes1g_{9/2})_p$ components couple in phase to the remainder parts in both states. 
Hence, the transition radii of the resulting full densities are similar. 
On the other hand, the 
($2d_{5/2}\otimes2d_{5/2})_n$ component couples in phase to the neutron-remainder transition densities in case of the 2$^+_1$ states and out of phase in case of the MSS.
Their coupling shifts $R_{\rm tr}$ of the 2$^+_1$ and $2^+_{\rm ms}$ states to larger and smaller $r$, respectively. 
Furthermore, the ($2d_{5/2}\otimes2d_{5/2})_n$ density peaks at larger $r$ than the remainder parts due to the additional node of the wave function.
Moreover, the matter transition radius depends on the relative amplitudes of proton and neutron densities. 
Thus, in-phase coupling increases $R_{\rm tr}$ of the corresponding 2$^+_1$ state, while out-of-phase coupling in case of the MSS has the opposite effect.

The differences seen between the matter transition radii of the $2^+_1$  and $2^+$ MSS are caused by two reasons. 
The first is a change of the neutron transition radii due to the peculiar density of the ($2d_{5/2}\otimes2d_{5/2})_n$ 2p state and the opposite  sign to the remainder parts. 
The second is the change of sign of the main neutron component which reduces the amplitudes of the full neutron transition density relative to the full proton transition density. 
Both effects contribute to the reduction of $R_{\rm m}$ of MSS with respect to the 2$^+_1$ state. 
The latter effect also explains why the matter transition radius of the 2$^+_{\rm ms}$ state of $^{94}$Zr is smaller than the one of the 2$^+_1$ state although its neutron transition radius is larger.

The theoretical analysis also provides a simple explanation for the different magnitudes and signs of $\Delta R_{\rm m}$ between $^{94}$Zr, $^{92}$Zr and $^{94}$Mo. 
The shift is strongest in $^{94}$Mo, where the neutron-remainder part relative to the main neutron component is weakest leading to strong cancellation effects for the MSS at larger radii. 
$^{92}$Zr is an intermediate case between $^{94}$Mo and $^{94}$Zr. 
In comparison to $^{94}$Mo, the remainder neutron part is stronger and additionally the contribution of the ($2d_{5/2}\otimes2d_{5/2})_n$-component to the wave function of the mixed-symmetry state is reduced from 44\% in $^{94}$Mo to 28\%. 
In $^{94}$Zr, this amplitude is further reduced to 20\%, and the remainder neutron part increases. 
Additionally, the large amplitude of ($2d_{5/2}\otimes3s_{1/2})_n$ 2qp state 
influences the neutron transition density of the MSS.

In view of these findings it would be interesting to measure the proton scattering cross section of the mixed-symmetry state of $^{96}$Mo. 
Here, the proton $B(E2)$ value of the MSS is 0.08$^{+0.02}_{-0.01}$ W.u.~\cite{lesher2007} only indicating a small contribution from high-lying 2qp states. 
Therefore, $\Delta R_{\rm m}$  and the corresponding shift in the 2$^+_1$ and 2$^+_{\rm ms}$ angular distributions might even be stronger than in $^{94}$Mo. Unfortunately, this state could not be identified in the experiments described in Paper I. 
This might be partly due to the above described cancellation effect, which not only reduces the neutron transition radius, but also the absolute proton scattering cross sections.

\begin{figure} 
\centering
%    \subfigure{\includegraphics[scale=0.4]{comparison2+_92zr.pdf}}\quad\quad
%    \subfigure{\includegraphics[scale=0.4]{comparison2+_94mo.pdf}}
\includegraphics[width=0.85\columnwidth]{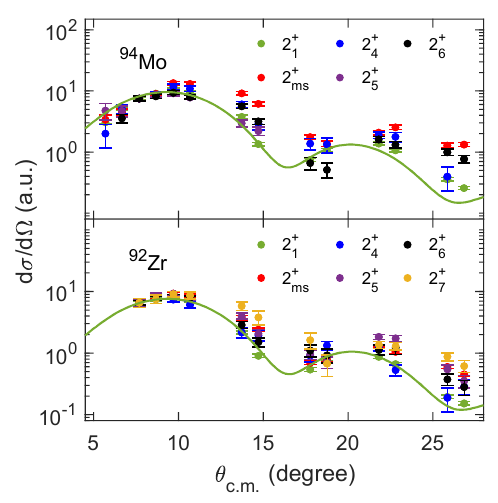}
    \caption[Comparison of the cross sections of all measured 2$^+$ states of $^{94}$Mo and $^{92}$Zr.]{\label{Fig: comparison2+_all}
     Comparison of the cross sections of all 2$^+$ states of $^{94}$Mo and $^{92}$Zr identified in Paper I.
     The cross sections are scaled to the one of the corresponding 2$^+_1$ states at $7.7^\circ$. 
     The solid green line is a theoretical 2$^+$ cross section calculated with a phenomenological optical potential and the collective model describing the 2$^+_1$ states. 
     }
\end{figure}

The QPM predicts the changes in the proton transition densities of the investigated nuclei to be small. However, one cannot exclude that part of the observed shift between the proton scattering cross sections of the 2$^+_1$ and 2$^+_{\rm ms}$ states is due to different proton transition radii. 
Electron scattering at low momentum transfers allows for verification of the QPM predictions. 
As shown in Fig.~\ref{Fig: ee comparison of quadrupole states}, the electron scattering form factors of $^{92}$Zr are similar in shape and are well described by the QPM calculations indicating similar proton transition radii. 
Data for $^{94}$Zr \cite{scheikh2014} are also consistent with comparable transition radii.

So far, 2$^+$ states other than 2$^+_1$ and 2$^+_{\rm ms}$ have been excluded from the discussion. 

This is justified because all other low-lying $2^+$ states are comparatively weakly excited from the ground state indicating small one-phonon components.   
The contribution of high-lying 2qp states is small and the shape of the proton scattering cross sections is determined by their main valence components. 
Hence, $R_{\rm m}$ can be different for each non-collective 2$^+$ state depending on the most important 2qp configuration. 

Figure~\ref{Fig: comparison2+_all} compares the angular distributions of all 2$^+$ states measured in Paper I for $^{94}$Mo and $^{92}$Zr. 
For $^{94}$Zr, no additional 2$^+$ states were observed. 
The cross sections are scaled to each other using the data points at the most forward angle. 
Clearly, the MSS of $^{94}$Mo is shifted to larger scattering angles than all other 2$^+$ states and thus has the smallest matter transition radius. 
In $^{92}$Zr, where the effect is weaker, the MSS has the second smallest matter transition radius. 
The 2$^+_7$ state with the smallest matter transition radius is at an energy of 3488~keV and thus far away from the expected energy region of the MSS.
In these two nuclei, the observed shift is strong enough to allow a differentiation between the MSS and all other 2$^+$ states. 
Thus, in nuclei with similar or larger $\Delta R_{\rm m}$ and no information on $B(M1)$ decays to the 2$^+_1$ state, one can identify the MSS by comparison of the proton scattering angular distributions populating the low-lying $2^+$ states.

\section{\label{sec:IV}Origin of low-energy collectivity}

In Ref.~\cite{walz2011}, a mechanism was discussed regarding how low-energy quadrupole states can gain collectivity from coupling to the ISGQR.
Here, an in-depth discussion is provided and the approach is extended to $3^-$ and $4^+$ one-phonon states including mixed-symmetry candidates.

\subsection{\label{sec:A}Quadrupole states}

\begin{figure} \centering
    \includegraphics[width=\columnwidth]{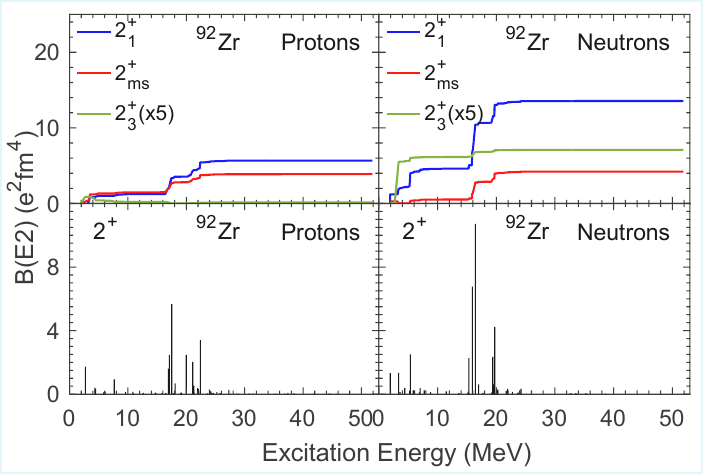}
    \caption[Running sums of the $B(E2)$ values of the lowest quadrupole QRPA phonons in $^{92}$Zr.]
    {\label{Fig: decomposition_2}
    Top: Running sums of the proton (left) and neutron (right) $B(E2)$ values for the [2$^+_1]$, [2$^+_{\rm ms}]$, and the noncollective [2$^+_3]$ states of $^{92}$Zr. 
    Bottom: Contributions of individual 2qp states.
    }
\end{figure}

Taking $^{92}$Zr as an example, Fig.~\ref{Fig: decomposition_2} shows the evolution of the $E2$ transition strengths to the ground state of the $2^+$ FSS and MSS calculated within the QPM as function of the excitation energy of the considered RPA components. 
In the top part, the running sums (similar to Fig.~\ref{Fig: ratio}) of the proton and neutron $B(E2)$ values are shown.
For comparison, the non-collective $2^+_3$ state is also presented.
A significant part of the $B(E2)$ strength of the collective states stems from 2qp states in the energy region 15-25 MeV where the ISGQR is located.
A similar effect on isotope shifts due to the coupling to giant resonances was discussed in Ref.~\cite{sagawa1987}.
The same high-lying states are responsible for the drop of $R(2^+_{\rm ms}$) from $\approx 1$ to $\approx 0$ in Fig.~\ref{Fig: ratio} precluding the use of Eq.~(\ref{eq: ratio}) as a signature of MSS.
In contrast, the non-collective $2^+_3$ state has small contributions from high-lying 2qp configurations. 
Here, $R \approx 1$ indicating a pure 2qp state.

At first glance, the large contribution of 2qp states at high excitation energies to the $B(E2)$ strengths comes as a surprise since their contribution to the QRPA norm of the wave functions is typically less than 1\%.
In the bottom row of Fig.~\ref{Fig: decomposition_2}, the proton and neutron $B(E2)$ values of each 2qp state are displayed. 
The contributions in the region 15-25 MeV are clearly enhanced in comparison to those of the valence space. The particle-hole residual interaction mixes them into the wave functions of the 2$^+_1$ and 2$^+_{\rm ms}$ states leading to $B(E2)$ strengths of several tens of W.u.\ for the $2^+_1$ and a few W.u.\ for the $2^+_{\rm ms}$ states. 
Hence, their small RPA amplitudes are compensated by large single-particle matrix elements.

\begin{figure} 
\centering
    \includegraphics[width=0.75\columnwidth]{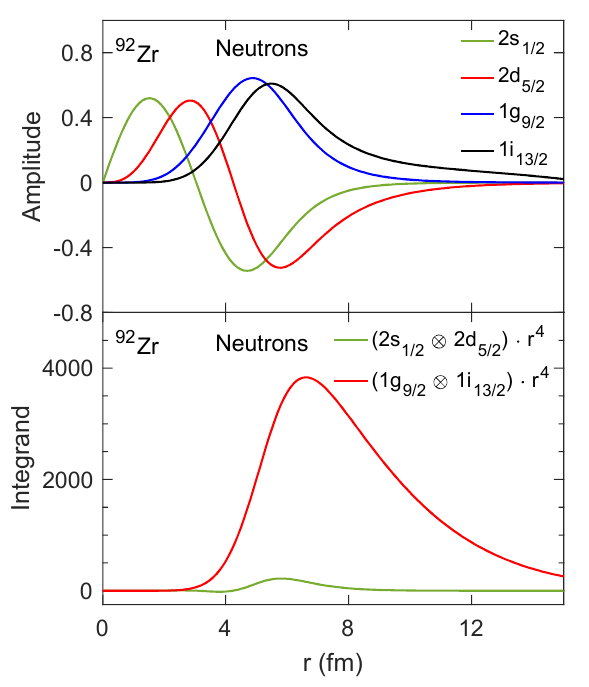}
    \caption[Radial wave functions of four neutron single-particle states.]
    {\label{Fig: wavefunctionsME} Top: Radial wave functions of four neutron single-particle states. Bottom: Radial dependence of the integrand of Eq.~(\ref{eq: matrixelement-coordinate-space}) for $\lambda$=2.}
\end {figure}

In order to understand why these matrix elements are so drastically enhanced, the top part of Fig.~\ref{Fig: wavefunctionsME} displays the radial parts of the $1g_{9/2}$ and $1i_{13/2}$ single-particle components that form the 2qp state ($1g_{9/2}\otimes1i_{13/2})_n$ with the largest individual $B(E2)$ values. 
For comparison, the radial $2s_{1/2}$ and $2d_{5/2}$ single-particle wave functions of the  ($2s_{1/2}\otimes2d_{5/2})_n$ state are also shown. 
Large $E2$ transition matrix elements are caused by a good spatial overlap of the two single-particle wave functions \cite{schiffer1976}
\begin{equation}
    \label{eq: matrixelement-coordinate-space}
    \langle j || E\lambda || j^{\prime} \rangle \sim \int^{\infty}_0 
    \varphi_j(r) \varphi_{j^{\prime}} r^{\lambda+2} dr.
\end{equation}
Here, $\varphi_i$ are the single-particle wave functions and $\lambda$ is the multipolarity of the transition. Due to the factor $r^{\lambda+2}$, a good overlap at larger distances to the center of mass of the nucleus is particularly important and, thus, favors single-particle states with large $l$ quantum numbers because of the centrifugal repulsion. 
The radial dependence of the integrand of Eq.~(\ref{eq: matrixelement-coordinate-space}) for both 2qp configurations is shown in the bottom part of Fig.~\ref{Fig: wavefunctionsME}. 
As expected, the ($1g_{9/2} \otimes 1i_{13/2}$) state is significantly enhanced compared to the ($2s_{1/2} \otimes 2d_{5/2}$) state in the surface region $5 -6$ fm.

The confinement of 2qp states with large contributions to the $B(E2)$ values of $2^+$ states to a 10~MeV wide excitation-energy window can be understood as follows: 
Besides the single-particle transition matrix element $\langle j || E\lambda || j^{\prime}\rangle$ discussed above, the Bogoliubov occupation probabilities $u$ and $v$ are important.
For 2qp transitions to make a contribution to the $B(E2)$ values, $u^{(+)}_{jj^{\prime}}=u_j v_j^{\prime}+u_j^{\prime} v_j$ should be large. 
Particle-hole transitions with $u_j\approx$1, $v_j\approx$0, $u_{j^{\prime}}\approx$0, $v_{j^{\prime}}\approx$1) are enhanced, while they are strongly suppressed for particle-particle transitions with $u_j\approx$1, $v_j\approx$0, $u_{j^{\prime}}\approx$1, $v_{j^{\prime}}\approx$0.
All single-particle states within one major shell have the same parity except for the high-spin unique parity single-particle state pushed down from the next oscillator shell because of the spin-orbit force. 
In order to create a 2qp state with positive parity, two single-particle states from the same harmonic oscillator shell (0$\hbar\omega$) or different by 2$\hbar\omega$ or 4$\hbar\omega$ are needed. 
1$\hbar\omega$ and 3$\hbar\omega$ transitions cannot form a 2qp state with positive parity except with the unique-parity state. 
Since these have large $j$ values, there are a small number of or no positive parity single-particle states with $\Delta(\hbar \omega) = \pm 1$ and sufficiently high $j$ to couple to a total angular momentum $J=2$. 

\subsection{\label{sec:B} Octupole and hexadecapole states}

\begin{figure} 
\centering
    \includegraphics[width=\columnwidth]{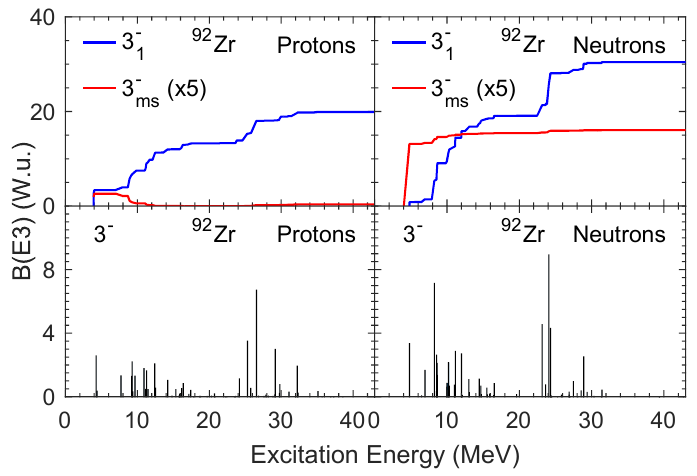}
    \caption[Running sums of the $B(E3)$-values of the lowest octupole QRPA-phonons in $^{92}$Zr.]
    {\label{Fig: decomposition_3} Same as Fig.~\ref{Fig: decomposition_2} but for $3^-$ states.}
\end{figure}

\begin{figure} 
\centering
    \includegraphics[width=\columnwidth]{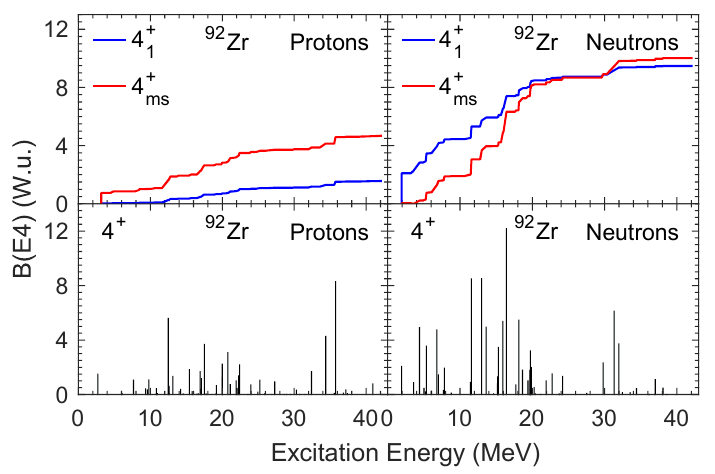}
    \caption[Running sums of the $B(E4)$-values of the lowest hexadecapole QRPA-phonons in $^{92}$Zr.]
    {\label{Fig: decomposition_4}
    Same as Fig.~\ref{Fig: decomposition_2} but for $4^+$ states.}
\end{figure}

Figures~\ref{Fig: decomposition_3} and \ref{Fig: decomposition_4} display the same kind of decomposition as in Fig.~\ref{Fig: decomposition_2} for 3$^-$ and 4$^+$ states, respectively.
Again, a large fraction of the $B(E\lambda$)
excitation strength from the ground state
of the [3$^-_1]$, [4$^+_1]$, and [4$^+_{\rm ms}]$ states is caused by 2qp states outside of the valence space. 
The valence space contributions to the transition strength of the $3^-_1$ phonon are very small since only configurations coupled to the unique-parity single-particle states can contribute.
The $3^-1$ phonon is non-collective, and its $B(E3)$ value is determined by the two main valence components. 

The contributions of individual 2qp states to the $B(E3)$ values are shown in the bottom part of Fig.~\ref{Fig: decomposition_3}. 
The single-particle states coupling to $J^{\pi}=3^-$ are 1$\hbar\omega$ or 3$\hbar\omega$ transitions explaining the two clusters of $B(E3)$ values at 5-15 MeV and 25-30 MeV, respectively.
The two excitation regions contribute about equally to the final collectivity of the low-energy octupole vibration.
While no collective octupole transitions are built in the $1\hbar\omega$ space, a part of the 2qp states forming the isoscalar giant dipole resonance at high excitation energies contribute. 

The running sums of the low-energy hexadecapole
strength from the ground state as function of the excitation energy of their RPA components in the top part of Fig.~\ref{Fig: decomposition_4} show a smooth increase up to an excitation energy of about 40 MeV.
The responsible 2qp states plotted in the bottom part are broadly distributed.
In particular, hexadecapole 2qp states also occur in the 1$\hbar\omega$ region. 
Unique-parity single-particle states such as the proton $1h_{11/2}$ and neutron $1i_{13/2}$ contribute to most of these 2qp states. 
Many of the single-particle states in the valence space can couple with them to $J=4$ in contrast to the octupole case discussed above. 
The shell structure of the nuclei considered here clearly supports the formation of collective one-phonon hexadecupole excitation. 
However, the residual interaction of higher multipoles ($J > 3$) is too weak to form collective structures at high excitation energies \cite{goeke1982}.

\section{\label{sec:V}Concluding remarks}

In this series of two papers, the identification of MSS in the vibrational  nuclei $^{92,94}$Zr and $^{94}$Mo and their specific properties based on a combined analysis of inelastic proton and electron scattering and pre-existing spectroscopic information are discussed. 
While Paper I is concerned with the data, Paper II presents a theoretical analysis of their wave functions based on the microscopic QPM. 
In order to evaluate its predictive power, an extensive comparison to the experimentally available information is performed including not only the proton and electron scattering cross sections but also g.s.\ properties, excitation energies, magnetic moments and electromagnetic transitions strengths.

The overall good description of the $2^+$ states permits the test of MSS signatures alternative to strong $M1$ transitions between MSS and FSS states. 
The ratio of proton and neutron transition matrix elements from the ground state [Eq.~(\ref{eq: ratio})] turns out to be not a good indicator because the collectivity of the quadrupole and hexadecupole FSS and MSS arises to a significant extent from coupling to high-lying states forming the ISGQR. 
However, the sign change between the leading proton and neutron 2qp configurations impacts the corresponding transition densities.
It can be decomposed by the combined analysis of inelastic proton and electron scattering and leads to a different momentum transfer dependence of the MSS compared to all other $2^+$ states in proton scattering.

WE also investigated the possible existence of octupole and hexadecapole MSS.
The QPM results identify MSS candidates with a sign change between the leading proton and neutron 2qp configurations compared to the FSS states.
However, in both cases the isoscalar quadrupole-quadrupole or quadrupole-octupole two-phonon states are predicted at comparable or even lower excitation energies than the MSS candidates, which implies considerable mixing and obscures an unambiguous identification. 
The $3^-$ MSS candidate is expected at much higher excitation energies than suggested in previous experiments because of the need to include the unique-parity single-particle orbital.
These findings indicate condierable fragmentation of the $3^-$ and $4^+$ MSS, which puts claims based on strong $M1$ transitions to the FSS into question.

Finally, the mechanism generating collectivity in the low-energy FSS and MSS is discussed.
For the quadrupole case, the $B(E2)$ strength of both are, to a large extent, generated by the strong coupling to high-lying $2\hbar\omega$ states forming the ISGQR.

For the octupole FSS, contributions from the 2qp states in the $1\hbar\omega$ and $3\hbar\omega$ shell regions are important, where a major part of the latter with different spin-couplings form the ISGDR.
However, the MSS candidate (based on the sign argument between the leading proton and neutron 2qp configurations) is predicted to be non-collective in the valence shell, i.e., with little contributions from high-energy 2qp components.
The collectivity of hexadecapole FSS and MSS is almost entirely generated by high-lying states broadly distributed between about 10 and 40 MeV. 

The new signature of $2^+$ MSS based on the difference of proton and neutron transition densities should work particularly well in cases with one or two bosons outside closed shells. 
It would be interesting to explore its range of applicability further.
For some of the interesting candidates ($^{70}$Zn,$^{96}$Mo) proton scattering data are already available, cf.\ paper I.

\section*{Acknowledgements}

We are indebted to J.~Wambach for many enlightening discussions on the topics of this paper.
This work was supported by the South African National Research Foundation (NRF) and by the Deutsche Forschungsgemeinschaft (DFG, German Research Foundation) under Grant No.\ SFB 1245 (project ID 279384907). 

\section*{Data Availability}

The data that support the findings of this article are not publicly available. 
They are available from the authors upon reasonable request.

\bibliography{MSSPaper2}

\end{document}